\documentclass[5p,twocolumn]{elsarticle}

\biboptions{sort&compress}

\usepackage{latexsym}
\usepackage{amsfonts}
\usepackage{color}
\usepackage{graphicx}
\usepackage{mathptmx}      
\usepackage{bm}
\usepackage{enumerate}
\usepackage{subfigure}
\usepackage{listings}
\usepackage{grffile}
\usepackage{multirow}
\usepackage{marvosym}
\usepackage{enumitem}
\usepackage{qcircuit}
\usepackage{float}
\usepackage{wrapfig}

\definecolor{DarkGreen}{rgb}{0.00,0.39,0.00}

\usepackage{qcircuit}

\newcommand\QC{{JUQCS}}
\newcommand\EQC{{JUQCS--E}}
\newcommand\AQC{{JUQCS--A}}
\newcommand{\onlinecite}[1]{\cite{#1}}
\def\KET#1{\vert #1 \rangle}
\def\mod{{\mathop{\hbox{ mod }}}}

\DeclareRobustCommand\openone{\leavevmode\hbox{\small1\normalsize\kern-.33em1}}

\journal{Computer Physics Communications}

\begin{document}

\begin{frontmatter}



\title{Massively parallel quantum computer simulator, eleven years later}


\cortext[cor1]{Corresponding author: s.yuan@whu.edu.cn}
\cortext[cor2]{Corresponding author: k.michielsen@fz-juelich.de}
\address[RUG]{Zernike Institute for Advanced Materials, University of Groningen, \\ Nijenborgh 4, NL-9747 AG Groningen, The Netherlands}
\address[FZJ]{Institute for Advanced Simulation, J\"ulich Supercomputing Centre, \\ Forschungzentrum J\"ulich, D-52425 J\"ulich, Germany}
\address[RWTH]{RWTH Aachen University, D-52056 Aachen, Germany}
\address[RIK]{RIKEN Center for Computational Science, 7-1-26 Minatojima-minami-machi,\\ Chuo-ku, Kobe, Hyogo 650-0047, Japan}
\address[TOK]{Department of Applied Physics, School of Engineering, The University of Tokyo, \\ Hongo 7-3-1, Bunkyo-ku, Tokyo 113-8656, Japan}
\address[WUH]{School of Physics and Technology, Wuhan University, \\Wuhan 430072, China }

\author[RUG]{Hans De Raedt}
\author[FZJ]{Fengping Jin}
\author[FZJ,RWTH]{Dennis Willsch}
\author[FZJ,RWTH]{Madita Willsch}
\author[RIK]{Naoki Yoshioka}
\author[RIK,TOK]{Nobuyasu Ito}
\author[WUH]{\\Shengjun Yuan\corref{cor1}}
\author[FZJ,RWTH]{Kristel Michielsen\corref{cor2}}

\begin{abstract}
A revised version of the massively parallel simulator of a universal quantum computer,
described in this journal eleven years ago,
is used to benchmark various gate-based quantum algorithms
on some of the most powerful supercomputers that exist today.
Adaptive encoding of the wave function reduces the memory requirement by a factor of eight, making it possible to simulate
universal quantum computers with up to 48 qubits on the Sunway TaihuLight and on the K computer.
The simulator exhibits close-to-ideal weak-scaling behavior on the Sunway TaihuLight,
on the K computer, on an IBM Blue Gene/Q, and on Intel Xeon based clusters, implying
that the combination of parallelization and hardware
can track the exponential scaling due to the increasing number of qubits.
Results of executing simple quantum circuits and
Shor's factorization algorithm on quantum computers containing up to 48 qubits are presented.
\end{abstract}

\begin{keyword} 
quantum computation\sep computer simulation\sep high performance computing\sep parallelization\sep  benchmarking
\PACS{03.67.Lx, 02.70.-c}
\end{keyword}
\date{\today}

\end{frontmatter}

\section{Introduction}\label{INTRO}

Simulating universal quantum computers on conventional, classical digital
computers is a great challenge. Increasing the number of qubits (denoted by $N$)  of the quantum
computer by one requires a doubling of the amount of memory of the digital
computer. For instance, to accurately simulate the operation of a universal
quantum computer with 45 qubits, one needs a digital computer with slightly more
than 1/2 Petabytes ($10^{15}/2$ bytes) of memory.  There are only a few digital
computers in the world which have the amount of memory, number of compute nodes,
and a sufficiently powerful network connecting all the compute nodes to perform
such simulations. Performing computations with such a large amount of memory and
processors requires a simulator that can efficiently use the parallel
architecture of present day supercomputers.

We report on novel algorithms and techniques implemented
in the J\"ulich universal quantum computer simulator (\QC).
In this paper, ``universal quantum computer'' refers to the theoretical, pen-and-paper, gate-based
model of a quantum computer~\cite{NIEL10} in which the time evolution of the machine is defined
in terms of a sequence of simple, sparse unitary matrices, with no reference to the real time
evolution of a physical system.
An article about an earlier version of the same simulator was published in this journal eleven years ago~\cite{RAED07x}.
Since then, supercomputer hardware has evolved significantly and therefore we thought it was time to
review and improve the computationally critical parts of the simulator and use it to
benchmark some of the most powerful supercomputers that are operational today.
In Table~\ref{SUPER} we collect the main characteristics of the computer systems
that we have used for our benchmarks.

\begin{table*}[t]
\caption{Overview of the computer systems used for benchmarking.
The IBM Blue Gene/Q JUQUEEN~\cite{JUQUEEN} (decommissioned), JURECA~\cite{JURECA}
and JUWELS are located at the J\"ulich Supercomputing Centre in Germany,
the K computer of the RIKEN Center for Computational Science in Kobe, Japan,
and the Sunway TaihuLight~\cite{SUNWAY} at the National Supercomputer Center in Wuxi, China.
The row ``\# qubits'' gives the maximum number of qubits $N$ that can be simulated with \AQC\ (\EQC).
At the time of running the benchmarks on JUWELS, the maximum number of qubits $N$ was limited to 43 (40).
}
\begin{center}
\begin{tabular}{cccccc}
\hline\noalign{\vskip 4pt}                                                                                %
                  & JUQUEEN       & K computer               & Sunway TaihuLight & JURECA-CLUSTER & JUWELS \\   
\hline\noalign{\vskip 4pt}                                                                                                 %
 CPU              & IBM PowerPC& eight-core SPARC64& SW26010 manycore            &  Intel Xeon    &  Dual Intel Xeon    \\   
                  & A2         & VIIIfx            & 64-bit RISC                 &  E5-2680 v3    &   Platinum 8168    \\   
\hline\noalign{\vskip 4pt}                                                                                       %
 clock frequency  &  1.6 GHz      & 2.0 Ghz                  & 1.45 GHz          &  2.5 GHz       &  2.7 GHz    \\   
 memory/node      &  16 GB                  & 16 GB          & 32 GB             &  128 GB        &  96 GB   \\           
 \# threads/core used  &  1 -- 2                 & 8              & 1                 &  1 -- 2        &  1 -- 2    \\   
 \# cores used    &  1 -- 262144            & 2 -- 65536     & 1 -- 131072       &  1 -- 6144     &  1 -- 98304  \\   
 \# nodes used    &  1 -- 16384             & 2 -- 65536     & 1 -- 32768        &  1 -- 256      &  1 -- 2048    \\   
 \# MPI processes used &  1 -- 524288            & 2 -- 65536     & 1 -- 131072  &  1 -- 1024     &  1 -- 2048    \\   
 \# qubits        &  46 (43)                & 48 (45)        & 48 (45)           &  43 (40)       &  46 (43)       \\   
\hline
\end{tabular}
\label{SUPER}
\end{center}
\end{table*}

\QC\ runs on digital computers ranging from personal computers
to the largest supercomputers that are available today.
The present version of the simulator comes in two flavors.
One version, referred to as \EQC\ (E referring to numerically exact, see below), uses double precision (8-byte) floating point arithmetic and
has been used to simulate a universal quantum computer with up to 45 qubits.
The 45 qubit limit is set by the amount of RAM memory available on the supercomputers that we have access to, see Table~\ref{SUPER}.
For a system of $N$ qubits and using 16 bytes per complex coefficient of the $2^N$ different basis
states, the amount of memory required to store the wave function is $2^{N+4}$, i.e., $1/2$ PB
are needed to store the wave function of $N=45$ qubits.
Adding storage for communication buffers (default is to use  $2^{N-3}$ bytes) and an insignificant
amount of bytes for the code itself, simulating a $N=45$ qubit universal quantum computer
requires a little more than $1/2$ PB but certainly less than  $1$ PB of RAM memory.

A second version, referred to as \AQC\ (A referring to approximate), trades memory for CPU time and
can be used to simulate a universal quantum computer with up to 48 qubits
on digital computers with less than 1 PB of RAM memory,
with a somewhat reduced numerical precision relative to the other version of the simulator.
\AQC\ employs adaptive coding to represent the quantum state in terms of 2-byte numbers,
effectively reducing the memory requirements by a factor of eight relative to the one of \EQC\ (see Sec.~\ref{AQS} for more details).
The adaptive coding requires additional computation such that for some of the quantum gates,
\AQC\ takes a longer time to complete than \EQC.
The reduced precision (about 3 digits) has been found more than sufficient
for all quantum circuits that have been tested so far.

From the quantum computer user perspective, \EQC\ and \AQC\ are fully compatible.
In this document, the acronym \QC\ refers to both versions while
\EQC\ and \AQC\ are used specifically to refer to the numerically exact version
and the adaptive-coding version of the simulator, respectively.
The only difference, if any, between \EQC\ and \AQC\ is in the accuracy of the results.

A quantum gate circuit for a universal quantum computer
is a representation of a sequence of matrix-vector operations involving matrices that are extremely sparse.
Only a few arithmetic operations are required to update one coefficient of the wave function.
Therefore, in practice, simulating universal quantum computers is rather simple
as long as there is no need to use distributed memory or many cores and the access
to the shared memory is sufficiently fast~\cite{RAED06,SMEL16,KHAM17,HANE17}.
The elapsed time it takes to perform such operations is mainly limited by the bandwidth to (cache) memory.
However, for a large number of qubits, the only viable way to alleviate the memory access problem
is to use distributed memory, which comes at the expense of overhead due to communication between nodes,
each of which can have several cores that share the memory (as is the case on all machines listed in Table~\ref{SUPER}).
Evidently, the key is to reduce this overhead by minimizing the transfer of data between nodes,
which is exactly what \QC\ does~\cite{RAED07x}.

Another road to circumvent the memory bottleneck is to use the well-known fact that
propagators involving two-body interactions (two qubits in the case at hand)
can be replaced by single-particle propagators by means of a Hubbard-Stratonovich transformation,
that is by introducing auxiliary fields.
A discrete version of this trick proved to be very useful in quantum Monte Carlo simulations of interacting fermions~\cite{HIRS83}.
In Sec.~\ref{section4}, we show that the same trick can be used in the present context to great advantage as well,
provided that the number of two-qubit gates is not too large
and that it is sufficient to compute only a small fraction of the matrix elements between basis states and the final state.
The latter condition considerably reduces the usefulness of this approach because for an algorithm such as Shor's,
it is a-priori unknown which of the basis states will be of interest.
Nevertheless, this trick of trading memory for CPU time is interesting in itself and has recently
been used, in various forms and apparently without recognizing the relation to the auxiliary field approach
to many-body physics, to simulate large random circuits with low depth~\cite{PEDN17,BOIX17,CHEN18,BOIX18}.

\QC\ is a revised and extended version of the simulator, written in Fortran, developed about eleven years ago~\cite{RAED07x}.
Depending on the hardware,
the source code can be compiled to make use of OpenMP, the Message Passing Interface (MPI), or a combination of both.
Apart from a few technical improvements, the ``complicated'' part of the simulator, i.e. the MPI communication scheme,
is based on the same approach as the one introduced eleven years ago~\cite{RAED07x}.
\EQC\ and \AQC\ use the same MPI communication scheme.
During the revision, we have taken the opportunity to add some new elementary operations for implementing error-correction schemes
and a translator that accepts circuits
expressed in OpenQASM, i.e., the language used by the IBM Q Experience~\cite{IBMQE,CROS17}.
The executable code of \QC\ has been built using a variety of
Fortran compilers such as Intel's ifort, GNU's gfortran, IBM's XLF, and others.
Using \AQC\ (\EQC), a notebook with 16GB of memory can readily simulate a universal quantum computer with 32 (29) qubits.
Since portability is an important design objective, we have not engaged in optimizing the code
on the level of machine-specific programming to make use of, e.g.,
the accelerator hardware in the Sunway TaihuLight.
We leave this endeavor to future work.

A \QC\ program looks very much like a conventional assembler program, a sequence of mnemonics with a short list of arguments.
\QC\ converts a quantum circuit into a form that is suitable as input for the simulation
of the real-time dynamics of physical qubit models, such as NMR quantum computing~\cite{RAED02}
using the massively-parallel quantum spin dynamics simulator SPI12MPI~\cite{RAED06},
or quantum computer hardware based on superconducting circuits~\cite{WILL17a}.
A description of the instruction set that \QC\ accepts is given in~\ref{APPb}.

The primary design objective of the original \QC\ simulator~\cite{RAED07x} was to provide
an environment for testing and optimizing the MPI communication part of SPI12MPI.
The efficient simulation of spin-1/2 models (e.g. physical models of quantum computers)
requires elementary operations that are significantly more complex than
those typically used in universal quantum computation~\cite{RAED06}.
Therefore, to test the MPI communication part properly, \QC\ does not exploit the special structure of the CNOT and the Toffoli gate
and also does not modify the input circuit using quantum gate circuit optimization techniques.

\QC\ is found to scale very well as a function of the number of compute nodes,
beating the exponential growth in time that is characteristic for simulating universal quantum computers~\cite{NIEL10}.
Such simulations can be very demanding in terms of processing power, memory usage, and network communication.
Therefore, \QC\ can also serve as a benchmark tool for supercomputers.
We cover this aspect by reporting weak scaling plots obtained by
running quantum algorithms on the supercomputers listed in Table~\ref{SUPER}.

The paper is structured as follows.
In Sec.~\ref{section2}, we briefly review the basics of gate-based universal quantum computing,
emphasizing the aspects which are important for the design of a simulator.
Section~\ref{section3} addresses techniques for distributing the workload
of a simulation over many compute cores.
The primary bottleneck of simulating a gate-based universal quantum computer is the amount
of memory required to store the wave function representing the quantum state of the machine.
Section~\ref{section4} discusses two very different methods for alleviating this problem.
In Sec.~\ref{section5}, we present results
obtained by executing a variety of quantum circuits on \QC, running
on five different supercomputers.
Conclusions are given in Sec.~\ref{section6}.

\section{Basic operation}\label{section2}

A quantum computer is, by definition, a device described by quantum theory.
The elementary storage unit of a quantum computer is, in its simplest form,
represented by a two-level system, called qubit~\cite{NIEL10}.
The state of the qubit is represented by a two-dimensional vector
\begin{equation}
|\Phi\rangle=a(0)|0\rangle+a(1)|1\rangle,
\label{STAT0}
\end{equation}
where $|0\rangle$ and $|1\rangle$ denote two orthogonal basis vectors of the
two-dimensional vector space and $a_0\equiv a(0)$ and $a_1\equiv a(1)$ are complex numbers,
normalized  such that $|a_0|^2+|a_1|^2=1$.

The internal state of a quantum computer comprising $N$ qubits is described by a $2^{N}$-dimensional unit vector
of complex numbers
\begin{eqnarray}
|\Phi\rangle&=&a({0\ldots00}) |0\ldots00\rangle
+a({0\ldots01}) |0\ldots01\rangle
+\ldots
\nonumber \\
&+&a({1\ldots10}) |1\ldots10\rangle
+a({1\ldots11}) |1\ldots11\rangle
,
\label{STAT2}
\end{eqnarray}
where
\begin{eqnarray}
\sum_{i=0}^{2^L-1}|a_i|^2=1
,
\label{STAT1}
\end{eqnarray}
i.e. by rescaling the complex-valued amplitudes $a_{i}$,
we normalize the vector $|\Phi\rangle$ such that $\langle\Phi|\Phi\rangle=1$.

Unlike in many-body physics where the leftmost (rightmost) bit of the basis state
represents quantum spin number 1 ($N$),
in the quantum computer literature it is common to label the qubits from 0 to $N-1$, that is
the rightmost (leftmost) bit corresponds to qubit 0 ($N-1$)~\cite{NIEL10}.

Executing a quantum algorithm on a universal, gate-based quantum computer
consists of performing a sequence of unitary operations on the vector $|\Phi\rangle$.
As an arbitrary unitary operation can be decomposed into a sequence of single-qubit operations
and the CNOT operation on two qubits~\cite{NIEL10},
it is sufficient to implement these specific operations
as a sequence of arithmetic operations on the vector
$\mathbf{v}=\left( a({0\ldots00}), \ldots, a({1\ldots11})\right)^\mathrm{T}$.

We illustrate the procedure for the Hadamard gate on qubit $0\le j \le N-1$.
The $2^{N}\times2^{N}$ matrix ${\cal H}$ multiplying the vector $\mathbf{v}$
is given by ${\cal H}=\openone_0\otimes\ldots\otimes\openone_{j-1}\otimes H\otimes
\openone_{j+1}\otimes\ldots\otimes\openone_{N-1}$
where $H$ denotes the $2\times2$ Hadamard matrix
\begin{equation}
H=\frac{1}{\sqrt{2}}\left(\begin{array}{rr} 1 & 1\\ 1 & -1 \end{array}\right)
.
\label{STAT5}
\end{equation}
Obviously, the matrix ${\cal H}$ is very sparse.
The matrix-vector multiplication $\mathbf{v} \leftarrow {\cal H}\mathbf{v}$
decomposes into $2^{N-1}$ matrix-vector multiplications involving $H$
and vectors of length two only.
The whole operation can be carried out in place (i.e. by overwriting $\mathbf{v}$),
requiring additional memory of ${\cal O}(1)$ only.

In more detail, the rule to update the amplitudes reads
\begin{eqnarray}
a({*\ldots*0_j*\ldots*})&\leftarrow&\frac{1}{\sqrt{2}}\bigg( a({*\ldots*0_j*\ldots*})
\nonumber \\
&&+a({*\ldots*1_j*\ldots*})\bigg)
\nonumber \\
a({*\ldots*1_j*\ldots*})&\leftarrow&\frac{1}{\sqrt{2}}\bigg( a({*\ldots*0_j*\ldots*})
\nonumber \\
&&-a({*\ldots*1_j*\ldots*})\bigg),
\label{NUM2}
\end{eqnarray}
where the $*$'s are placeholders for the bits $0,\ldots,j-1$ and $j+1,\ldots,N-1$.
From Eq.~(\ref{NUM2}), it follows that the update process consists
of selected pair of amplitudes using bit $j$ as index, replace the two amplitudes
by the right-hand side of Eq.~(\ref{NUM2}), and repeat the calculation for all possible pairs.
Obviously, the update process exhibits a very high degree of intrinsic parallelism.

Implementing quantum gates involving two qubits, e.g. the CNOT gate, or three qubits, e.g. the Toffoli gate,
requires selecting groups of four or eight amplitudes, respectively.
In other words, instead of the two rules in Eq.~(\ref{NUM2}), we have $2^{N-2}$ ($2^{N-3}$)
groups of four (eight) amplitudes that need to be updated.
Note that as the CNOT (Toffoli) gate only exchanges two of the four (eight) amplitudes,
the computational work involved is less than in the case of say the Hadamard gate.

The result of executing a quantum gate circuit on \QC\ is
an array of amplitudes $\mathbf{v}$ which represents the final state $|\Phi\rangle$ of the pen-and-paper quantum computer.
According to quantum theory, measuring a single qubit $j$
yields an outcome that is either 0 or 1 with probability
$\sum_{*} |a({*\ldots*0_j*\ldots*})|^2$
or
$\sum_{*} |a({*\ldots*1_j*\ldots*})|^2$, respectively.
\QC\ provides methods for computing these probabilities as well as for generating events.

\section{Parallelization techniques}\label{section3}

As explained in Sec.~\ref{section2}, the sparse matrix structure of the quantum gates
translates into an algorithm for updating the amplitudes which exhibits a very high degree of intrinsic parallelism.
In this section, we discuss two different techniques to exploit this parallelism.

\subsection{OpenMP}

Assuming memory is not an issue, a platform independent method to distribute the computational work over several compute cores
is to use OpenMP directives.
Thereby, care has to be taken that the order in which the groups of amplitudes are processed is ``cache friendly''.
The excerpt of the Fortran code given below shows how we have implemented a single-qubit gate in qubit $j$.

\smallskip
\begin{verbatim}
        nstates=2**(N-1)
        i=2**j
        if( nstates/(i+i) >= i ) then
!$OMP parallel do private ...
        do k=0,nstates-1,i+i
        do l=0,i-1
        i0 = ior(l, k)  ! *...* 0 *... *
        i1 = ior(i0, i) ! *...* 1 *... *
        ...
        end do
        end do
!$OMP end parallel do
        else
        do k=0,nstates-1,i+i
!$OMP parallel do private ...
        do l=0,i-1
        ...
        enddo
!$OMP end parallel do
        enddo
\end{verbatim}
%
\medskip

The indices k and l run over all possible values of the bits $(N-1),\ldots,j+1$ and $(j-1),\ldots,0$, respectively.
We use the test ``nstates/(i+i) $\ge$ i'' to decide whether it is more efficient to distribute both the outer and inner
loop or only the inner loop over all compute cores.
Our numerical experiments show that the test ``nstates/(i+i) $\ge$ i'' is not optimal.
The best choice depends on $N$ and on the particular hardware in a seemingly complicated manner
but the reduction of computation time is marginal.
Therefore, we opted for the simple, universal test ``nstates/(i+i) $\ge$ i''.

The implementation of two-qubit gates involving qubits $j_0$  and $j_1$ requires three instead of two loops to generate
the ``*'''s in the bit string ``$* \ldots* j_1 * \ldots* j_0 * \ldots *$'', a simple generalization
of the code used to implement single-qubit operations. This scheme straightforwardly extends to three-qubit gates.

\subsection{MPI}

As the number of qubits $N$ increases, there is a point at which a single compute node does not have enough memory to
store the whole vector of amplitudes such that it become necessary to use memory distributed over several compute nodes.
A simple scheme to distribute the amplitudes over a number of nodes is to use the high-order bits as
the integer representation of the index of the node.
Let us denote the number of high-order bits that will be used for this purpose by $N_{\mathrm{h}}$,
the corresponding number of compute nodes by $K_{\mathrm{h}}=2^{N_{\mathrm{h}}}$,
and the number of amplitudes per compute node by $K_{\mathrm{l}}=2^{N_{\mathrm{l}}}$
where $N_{\mathrm{l}}=N-N_{\mathrm{h}}$.

Obviously, there is no need to exchange data between compute nodes if we perform a single-qubit operation
on qubit $0\le j< N_{\mathrm{l}}$ because each pair of amplitudes which need to be updated
resides in the memory of the same compute node.
However, if $N_{\mathrm{l}}\le j< N-1$, it is necessary to exchange
data between compute nodes before the two amplitudes can be multiplied by the $2\times 2$ matrix
which represents the quantum gate.
For an operation such as the Hadamard gate, this implies that half of all amplitudes have to be
transferred to another compute node.
In \QC\ this exchange is implemented by swapping nonlocal qubits and local ones and keeping
track of these swaps by updating the permutation of the $N$ bit indices~\cite{RAED07x}.
Nevertheless, if $N$ is large, this swapping is a time-consuming operation,
even if the inter-node communication network is very fast.

Operations such as the CNOT and Toffoli gate that only exchange two amplitudes can be implemented without having to exchange
data through the inter-node communication network.
As explained earlier, the primary design objective of the original \QC\ simulator~\cite{RAED07x} was to provide
an environment for testing and optimizing the MPI communication scheme for a
quantum spin dynamics simulator which requires the implementation of more complicated many-qubit gates.
Therefore, the current version of \QC\ does not exploit the special structure of the CNOT or Toffoli gate.
The MPI communication scheme that we use is, apart from its actual implementation, identical
to the one described in Ref.~\cite{RAED07x} and will therefore not be discussed in detail here.

\section{Trading memory for CPU time}\label{section4}

The main factor limiting the size of the pen-and-paper quantum computer that can be simulated
is the memory required to store the $2^N$ amplitudes of the vector $|\Phi\rangle$.
In this section, we discuss two different methods to reduce the amount of memory needed.
Evidently, this reduction comes at the price of an increase of computation time.

\subsection{Double precision versus byte encoding}\label{AQS}

In quantum theory, the state of a single qubit is represented by
two complex numbers $\psi_0$ and $\psi_1$ which are normalized such that $|\psi_0|^2+ |\psi_1|^2 =1$~\cite{NIEL10}.
A gate operation on the qubit changes these numbers according to
\begin{eqnarray}
\left(
\begin{array}{c}
\psi_0\\
\psi_1\\
\end{array}
\right)
\leftarrow
U
\left(
\begin{array}{c}
\psi_0\\
\psi_1\\
\end{array}
\right)
,
\label{AQS0}
\end{eqnarray}
where $U$ is a $2\times2$ unitary matrix.
Gate operations involving $n$ qubits correspond to (repeated) matrix-vector multiplications
involving $2^n\times2^n$ unitary matrices.
As the number of arithmetic operations on the vector of complex amplitudes representing the state
of the $N$-qubit systems grows exponentially with $N$, it may seem necessary
to perform these operations with high numerical precision.
Our implementation of \EQC\ uses two 8-byte floating point numbers to encode one complex amplitude.

However, not all gates change the numerical representation of the state amplitudes.
For instance, the X and CNOT gates only swap amplitudes,
while the Hadamard gate arithmetically combines the two amplitudes.
Therefore, we have explored various ways to encode the complex numbers with less than 16 bytes.
An adaptive encoding scheme that we have found to perform quite well for quantum gate circuits
is based on the polar representation
$z=r e^{i\theta}$ of the complex number $z$.
We use one byte variable $-128 \le b_1<128$ to encode the angle $-\pi\le\theta<\pi$, i.e. $\theta=\pi b_1/128$.
Another byte variable $-128 \le b_0<128$ is used to represent $r$ in the following manner.
The special values $r=0$ and $r=1$ correspond to $b_0=-128$ and $b_0=127$, respectively.
The remaining values of $-127 \le b_0\le126$ are used to compute $r$ according to
$r=(b_0+127)(r_1-r_0)/253+r_0$, where $r_0$ and $r_1$ are the minimum and maximum value
of the $z$'s with $0<|z|<1$ over all elements of the state vector.
The values of $r_0$ and $r_1$ need to be updated
to adaptively tune the encoding scheme to the particular quantum circuit being executed.
Obviously, our encoding scheme reduces the amount of memory required
to store the state by a factor of 8 at the expense of additional CPU time to perform the decoding-encoding procedure.
The amount of additional CPU time depends on the gate and varies from very little for e.g. the X or CNOT gate to
a factor of 3--4 for gates such as the Hadamard or +X gate.

\subsection{Auxiliary variable method}\label{AUX}
An appealing feature of the universal quantum computation model is that
only a few single-qubit gates and the CNOT gate suffice to
perform an arbitrary quantum computation~\cite{NIEL10}.
In other words, in principle, any unitary matrix can be written
as a product of unitary matrices that involve only single-qubit
and two-qubit operations.

This subsection demonstrates that any circuit involving single-qubit gates, controlled-phase-shifts, and CNOT
operations can be expressed as a string of single-qubit operations, summed over a set of discrete, two-valued auxiliary variables.
Each term in this sum can be computed in ${\cal O}(N)$ arithmetic operations.
The number of auxiliary variables is exactly the same as the number $P$ of controlled-phase-shifts or CNOT gates in the circuit.
The worst case run time and memory usage of this algorithm are ${\cal O}(N M 2^P)$
and ${\cal O}(N + M)$, respectively, where $M$ is the number of output amplitudes desired.
Clearly, if $M\ll 2^N$, the memory reduction from ${\cal O}(2^N)$ to ${\cal O}(N+M)$ bytes
becomes very significant as the number of qubits $N$ increases.
To be effective, this approach requires that the input state to the circuit is a product state.
However, this is hardly an obstacle because in the gate-based model of quantum computation, it is
standard to assume that the $N$-qubit device can be prepared in the product state~\cite{NIEL10}
\begin{eqnarray}
|0\rangle=|0\rangle_0|0\rangle_1 \ldots |0\rangle_{N-1}
,
\label{appa0}
\end{eqnarray}
where the subscripts refer to the individual qubits.

First, let us consider a string of single-qubit gates acting on qubit $j=0,\ldots,N-1$
and denote the product of all the unitary matrices corresponding to these single-qubit gates by
$V_j$.
The application of these gates changes the initial state of qubit 0 into
\begin{eqnarray}
V_j|0\rangle_j=\alpha_j|0\rangle_j + \beta_j|1\rangle_j
,
\label{appa1}
\end{eqnarray}
where $\alpha_j$ and $\beta_j$ are complex-valued numbers satisfying $|\alpha_j|^2+|\beta_j|^2=1$.
If $V=V_0\otimes\ldots\otimes V_{N-1}$
represents a circuit that consists of single-qubit gates only, we have
\begin{eqnarray}
V|0\rangle=\prod_{j=0}^{N-1}\left(\alpha_j|0\rangle_j + \beta_j|1\rangle_j\right)
.
\label{appa2}
\end{eqnarray}
From Eq.~(\ref{appa2}), it follows immediately that in practice, the right-hand side can be computed
in ${\cal O}(N)$ arithmetic operations on a digital computer.
More importantly, the amount of memory required to store the product state Eq.~(\ref{appa2}) is
only $2^5N$ bytes (assuming 8-byte floating point arithmetic), much less (if $N>7$) than the exponentially
growing number $2^{N+4}$ required to store an arbitrary state.

Second, consider the results of applying to the state Eq.~(\ref{appa2}), a CNOT gate with control qubit 0
and target qubit 1.
We have
\begin{eqnarray}
\mathrm{CNOT}_{01}V|0\rangle&=&
\bigg(\alpha_0\alpha_1|0\rangle_0|0\rangle_1 +\beta_0\alpha_1|1\rangle_0|1\rangle_1
\nonumber \\
&&+\alpha_0\beta_1|0\rangle_0|1\rangle_1 +\beta_0\beta_1|1\rangle_0|0\rangle_1 \bigg)
\nonumber \\
&&\times\bigg(\prod_{j=2}^{N-1}\left(\alpha_j|0\rangle_j + \beta_j|1\rangle_j\right)\bigg)
,
\label{appa3}
\end{eqnarray}
such that it is no longer possible to treat the coefficients of qubit 0 and 1 independently from each other.
Of course, this is just a restatement, in computational terms, that the CNOT gate is a so-called ``entangling'' gate.
It is not difficult to imagine that a circuit containing several CNOT (or controlled phase shift, Toffoli) gates that
involve different qubits can create a state, such as the one created by the sequence of CNOT gates mentioned
in Sec.~\ref{sectionCNOT}, in which a single-qubit operation on one
particular qubit changes the amplitudes of all basis states.
Thus, any strategy to reduce the memory usage must deal with this aspect and must therefore ``eliminate''
the entangling gates.

A simple, effective method to express controlled phase shifts and CNOT gates in terms of
single-qubit gates is to make use of the discrete Hubbard-Stratonovich transformation,
originally introduced to perform quantum Monte Carlo simulations of the Hubbard model~\cite{HIRS83}.
Consider the controlled phase shift operation defined by the unitary matrix
\begin{eqnarray}
U_{01}(a)=\left(\begin{array}{cccc}
1 & 0 & 0 & 0\\
0 & 1 & 0 & 0\\
0 & 0 & 1 & 0\\
0 & 0 & 0 & e^{ia}
\end{array}\right)
=e^{ia (1+\sigma^z_0\sigma^z_1-\sigma^z_0 - \sigma^z_1)/4}
,
\label{appa4}
\end{eqnarray}
where $\sigma^z_0$ and $\sigma^z_1$ are the $z$-components of the Pauli matrices representing qubit 0 and 1, respectively.
Note that by convention, the computational basis is built from eigenstates of the $z$-components of the Pauli matrices~\cite{NIEL10}.
When $a$ takes values $\pm 2\pi/2^k$, the matrix in Eq.~(\ref{appa4}) is exactly the one which performs the conditional phase
shifts in the quantum Fourier transform circuit;
and when $a=\pi$, we have $H_1 U_{01}(\pi) H_1 = \mathrm{CNOT}_{01}$ such that
all CNOT gates can be expressed as a product of Hadamard gates and $U_{01}(\pi)$.
Thus, without loss of generality, it is sufficient to consider $U_{01}(a)$ only.

Denoting the eigenvalues of $\sigma^z_0$ and $\sigma^z_1$ by $\sigma_0$ and $\sigma_1$, respectively,
we note that $e^{ia \sigma_0\sigma_1/4}$ can only take two values and can be written as
\begin{eqnarray}
e^{ia \sigma_0\sigma_1/4}=\frac{e^{-ia/4 }}{2}\sum_{s=\pm1}e^{ix (\sigma_0+\sigma_1)s}
\quad,\quad \sigma_0,\sigma_1=\pm1,
\label{appa5}
\end{eqnarray}
where $x$ is given by $\cos 2x=e^{ia/2}$.
Therefore, we have
\begin{eqnarray}
U_{01}(a)
&=&e^{ia (1+\sigma^z_0\sigma^z_1-\sigma^z_0 - \sigma^z_1)/4}
\nonumber \\
&=&\frac{1}{2}\sum_{s=\pm1}e^{i(\sigma^z_0+\sigma^z_1)(xs-a/4)}
,
\label{appa6}
\end{eqnarray}
and we have accomplished the task of writing the controlled phase shift Eq.~(\ref{appa4})
as a sum of products of two single-qubit operations each.

The final step is to introduce auxiliary variables $s_p=\pm1$ for
each of the $p=1,...,P$ controlled phase shifts (including those that originate from rewriting the CNOTs)
that appear in the quantum circuit.
Then, the result of applying the whole circuit to the initial state $|0\rangle$ can be written as
\begin{eqnarray}
|\psi\rangle=\frac{1}{2^P}\sum_{s_1\ldots s_P=\pm1} \prod_{j=1}^N W_j(s_1,\ldots,s_P)|0\rangle
,
\label{appa7}
\end{eqnarray}
where $W_j(s_1,\ldots,s_p)$ is a concatenation of single-qubit operations on qubit $j$.
The action of $W_j(s_1,\ldots,s_p)$ can be computed independently (and in parallel if desired)
of the action of all other $W_{j'}(s_1,\ldots,s_p)$'s.
In practice, for large $N$, the advantage of the auxiliary variable approach in terms of
memory usage disappears if the application requires knowledge of the full state
$|\psi\rangle$ but can be very substantial if knowledge of only a few of the $2^N$ amplitudes of $|\psi\rangle$ suffices.

\section{Validation and benchmarking}\label{section5}

\begin{table*}[t]
\caption{The expectation values of the individual qubits, measured after
performing a Hadamard operation on each of the $N$ qubits as obtained by \AQC\ and \EQC.
Recall that \AQC\ uses a factor of 8 less memory than \EQC\
but still yields the same numerically exact results as those produced by \EQC\ for these tests.
The \AQC\ calculations were performed on JUQUEEN (up to $N=46$ qubits),
Sunway TaihuLight (up to $N=48$ qubits),
the K computer (up to $N=48$ qubits), and JURECA (up to $N=43$ qubits).
The \EQC\ calculations were performed on JUQUEEN (up to $N=43$ qubits),
Sunway TaihuLight (up to $N=45$ qubits),
the K computer (up to 45 qubits),
JURECA (up to $N=40$ qubits),
and JUWELS (up to $N=40$ qubits).
The line beginning with `$\ldots$' is a placeholder for the results
of measuring qubits $1,\ldots,N-2$.
}
\begin{center}
\begin{tabular}{c|ccc|ccc}
\hline
     &\multicolumn{3}{c|}{\AQC}&\multicolumn{3}{c}{\EQC}\\
qubit&$\langle Qx(i) \rangle$&$\langle Qy(i) \rangle$&$\langle Qz(i) \rangle$&
$\langle Qx(i) \rangle$&$\langle Qy(i) \rangle$&$\langle Qz(i) \rangle$ \\
\hline
    0 & 0.000 & 0.500 & 0.500  & 0.000 & 0.500  &  0.500 \\
    $\ldots$ & 0.000 & 0.500 & 0.500  & 0.000 & 0.500  &  0.500 \\
   $N-1$ & 0.000 & 0.500 & 0.500  & 0.000 & 0.500  &  0.500 \\
\hline
\end{tabular}
\label{Htab}
\end{center}
\end{table*}

The first step in validating the operation of \QC\ is to
execute all kinds of quantum circuits, including circuits randomly generated
from the set of all the gates in the instruction set, for a small ($N=2$) to moderate ($N\approx30$)
number of qubits on PCs running Windows (7,10) and on Linux workstations.
Validating the operation of \QC\ when it makes use of MPI, OpenMP or both is less trivial, in particular
if the number of qubits is close to the limit of what can be simulated on a particular hardware platform.
Of course, validation of real quantum computing devices is much more difficult.
In contrast to a simulation on a digital computer where the full state of the quantum computer
is known with high accuracy, the correct operation of real quantum computing devices
must be inferred by sampling the amplitudes of the computational basis states,
a daunting task if the number of qubits increases.

On a PC/workstation with, say 16 GB of memory, one can run small problems,
i.e. those that involve not more than 29 (32 when \AQC\ is used) qubits.
Quantum circuits involving 45 or more qubits can only be tested on supercomputers
such as the IBM Blue Gene/Q of the J\"ulich Supercomputing Centre in Germany,
the K computer of the RIKEN Center for Computational Science in Kobe, Japan, or
the Sunway TaihuLight at the National Supercomputer Center in Wuxi in China.

Validating the operation of \QC\ requires circuits for which the exact input-output relation is known
such that the correctness of the outcome can be easily verified.
This section presents \QC\ results obtained by executing quantum circuits for which this is the case and,
at the same time, illustrates the scaling and performance of \QC\ on the supercomputers listed in Table~\ref{SUPER}.

\subsection{Uniform superposition}
A common first step of a gate-based quantum algorithm is to turn the initial state (all qubits in state $|0\rangle$) into
a uniform superposition by a sequence of Hadamard operations.
Such a sequence has the nice feature that it can be trivially extended to more and more qubits
and is therefore well-suited to test the weak scaling behavior of a universal quantum computer simulator.
Note that independent of the number of qubits and computer architecture used, it is possible
to construct the uniform superposition without any form of communication between nodes
(a technique used by the SHORBOX instruction, see Sec.~\ref{SHOR})
but, as explained in Sec.~\ref{INTRO}, one of the design objectives of \QC\ was to test and benchmark the MPI communication,
not to construct the most efficient simulator of a universal quantum computer tuned to specific hardware.
Therefore, on purpose, we do not ``optimize'' the quantum circuit at this level.

\begin{figure}[ht]
\begin{center}
\includegraphics[width=0.85\hsize]{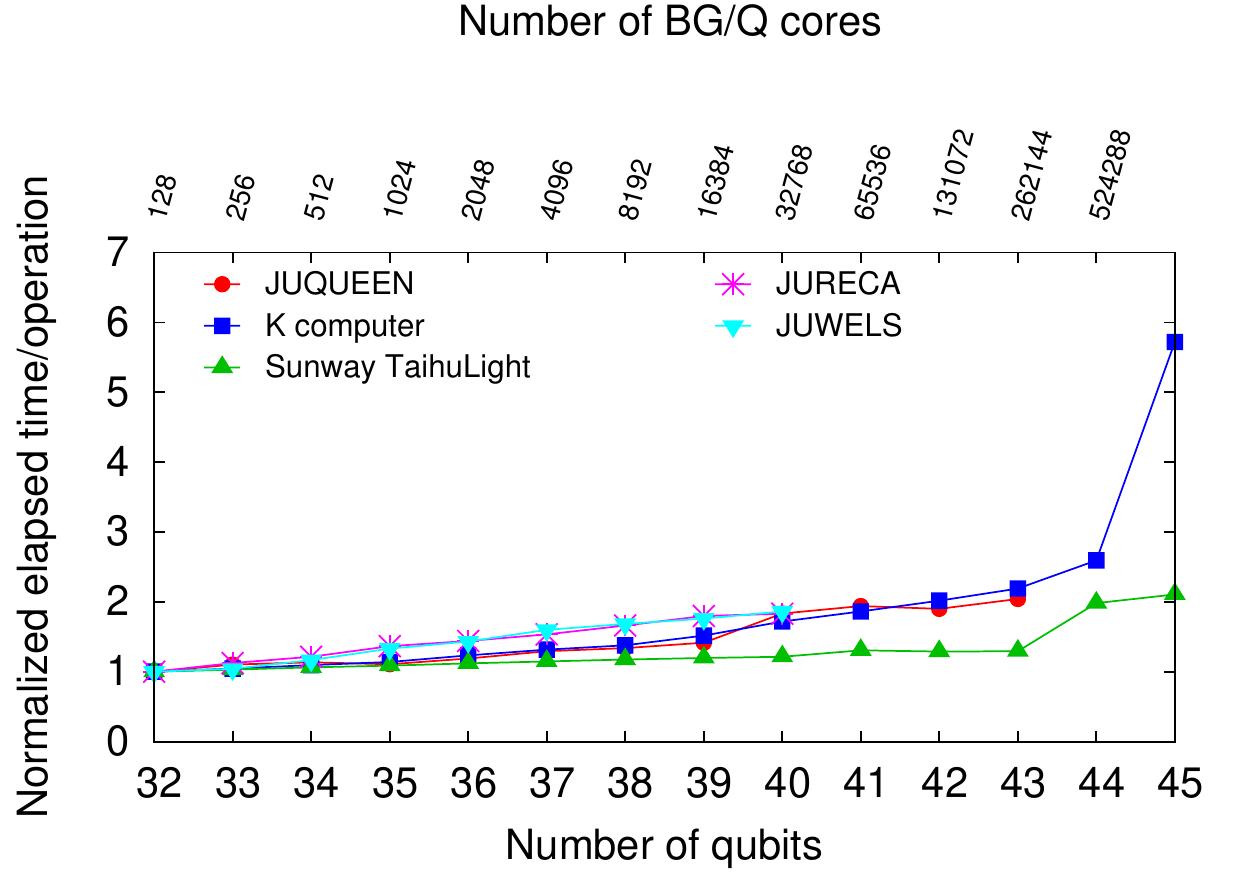}
\end{center}
\caption{The elapsed time per gate operation
(normalized by the values 1.2 s (JUQUEEN), 1.0 s (K), 7.7 s (Sunway TaihuLight), 1.9 s (JURECA),
and 1.3 s (JUWELS) for running the 32 qubit circuit)
as a function of the number of qubits, as obtained by \EQC\ executing
a Hadamard operation on each qubit. 
This weak scaling plot shows that \EQC\ beats the exponential
scaling of the computational work by doubling the size of the machine with each added qubit.
}
\label{scaling1a}
\end{figure}

\begin{figure}[ht]
\begin{center}
\includegraphics[width=0.85\hsize]{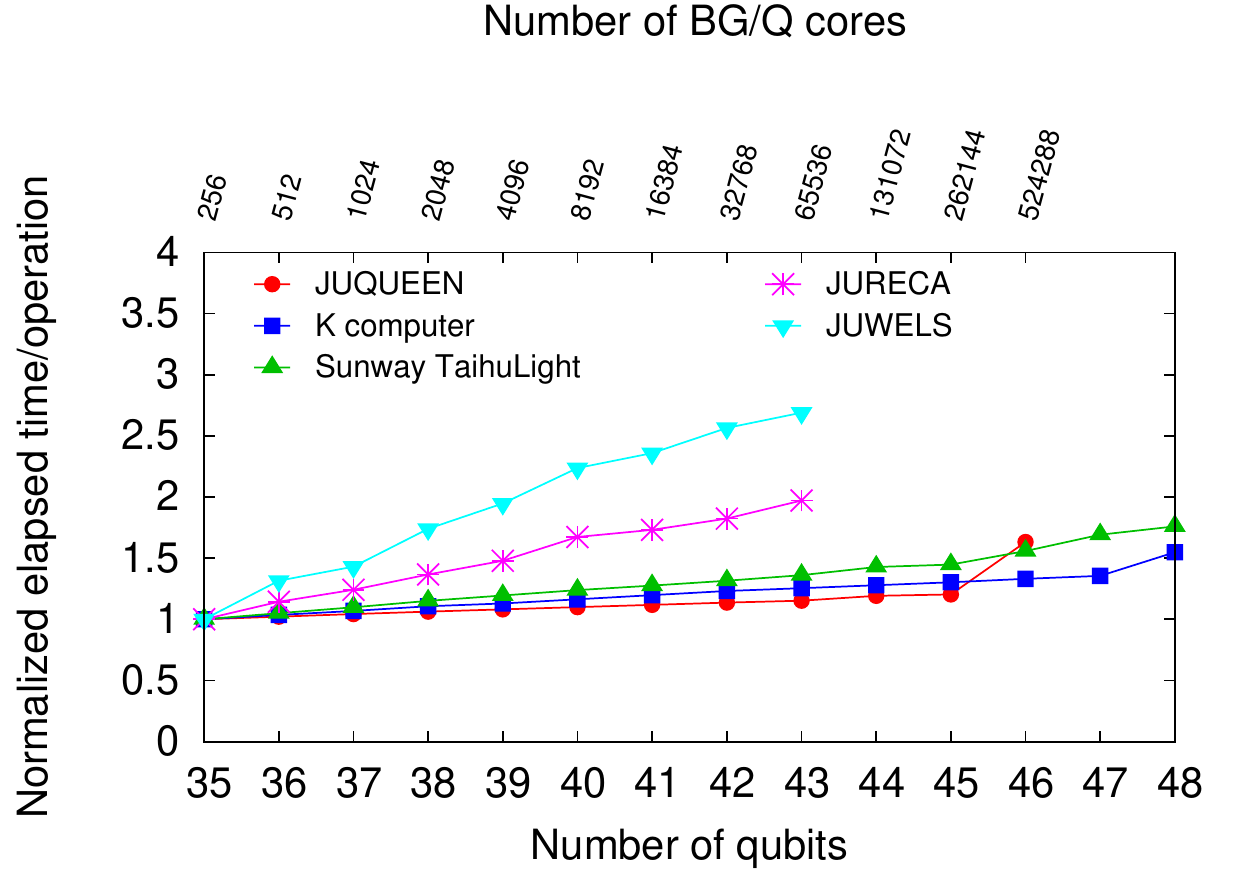}
\end{center}
\caption{The elapsed time per gate operation
(normalized by the values 10.7 s (JUQUEEN), 30.8 s (K), 101.2 s (Sunway TaihuLight),  9.7 s (JURECA)
and 4.7 s (JUWELS) for running the 35 qubit circuit)
as a function of the number of qubits, as obtained by \AQC\ executing
a Hadamard operation on each qubit. 
This weak scaling plot shows that \AQC\ beats the exponential
scaling of the computational work by doubling the size of the machine with each qubit added.
Note that  \AQC\ not only uses a factor of 8 less memory than \EQC\ but also uses
a factor of 8 less cores to run the same circuit.
}
\label{scaling1b}
\end{figure}

Table~\ref{Htab} summarizes the results of executing such sequences of Hadamard operations
on \EQC\ for $N\le45$ and on \AQC\ for $N\le48$.

In Figs.~\ref{scaling1a} and ~\ref{scaling1b} we present the results of a weak scaling analysis
of the elapsed times required to execute a Hadamard operation on each of the $N$ qubits.
Clearly, by doubling the size of the machine with each added qubit, \QC\ beats
the exponential scaling of the computational work with the number of qubits.

\begin{table*}[ht]
\caption{The expectation values of the individual qubits, measured after
performing the sequence (H 0), (CNOT 0 1), (CNOT 1 2), ..., (CNOT N-2, N-1),
followed by a measurement of all $N$ qubits.
Recall that \AQC\ uses a factor of 8 less memory than \EQC\
but, for these tests, also yields the same numerically exact results as those produced by \EQC.
The \AQC\ calculations were performed on JUQUEEN (up to $N=46$ qubits),
Sunway TaihuLight (up to $N=48$ qubits),
the K computer (up to $N=48$ qubits),
and JURECA (up to $N=43$ qubits).
The \EQC\ calculations were performed on JUQUEEN (up to $N=43$ qubits),
Sunway TaihuLight (up to $N=45$ qubits),
the K computer (up to 45 qubits),
JURECA (up to $N=40$ qubits),
and JUWELS (up to $N=40$ qubits).
The line beginning with `$\ldots$' is a place holder for the results of measuring qubits $1,\ldots,N-2$.
}
\begin{center}
\begin{tabular}{c|ccc|ccc}
\hline
     &\multicolumn{3}{c|}{\AQC}&\multicolumn{3}{c}{\EQC}\\
qubit&$\langle Qx(i) \rangle$&$\langle Qy(i) \rangle$&$\langle Qz(i) \rangle$&
$\langle Qx(i) \rangle$&$\langle Qy(i) \rangle$&$\langle Qz(i) \rangle$ \\
\hline
    0 & 0.500 & 0.500 & 0.500  & 0.500 & 0.500  &  0.500 \\
    $\ldots$ & 0.500 & 0.500 & 0.500  & 0.500 & 0.500  &  0.500 \\
   $N-1$ & 0.500 & 0.500 & 0.500  & 0.500 & 0.500  &  0.500 \\
\hline
\end{tabular}
\label{CNOTtab}
\end{center}
\end{table*}

\subsection{Sequence of CNOT gates}\label{sectionCNOT}
\begin{figure}[ht]
\begin{center}
\includegraphics[width=0.85\hsize]{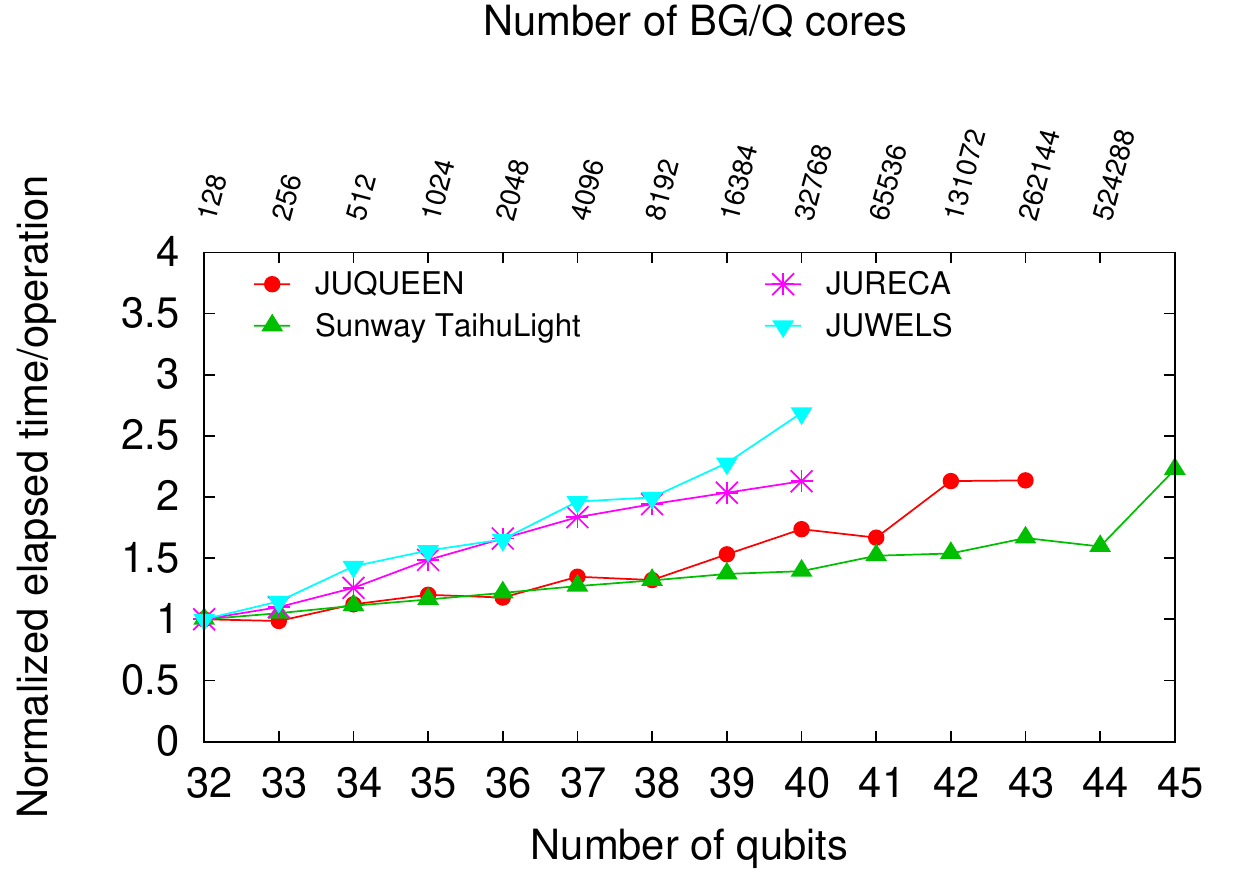}
\end{center}
\caption{The elapsed time per gate operation
(normalized by the values 0.98 s (JUQUEEN), 5.1 s (Sunway TaihuLight), 1.4 s (JURECA), and 0.9 s (JUWELS)
to run the 32 qubits circuit)
as a function of the number of qubits,
as obtained by \EQC\ executing a Hadamard operation on qubit 0 and the sequence
(CNOT 0 1), (CNOT 1 2), ..., (CNOT N-2, N-1), followed by a measurement
of the expectation values of all the qubits.
This weak scaling plot shows that \EQC\ beats the exponential
scaling of the computational work by doubling the size of the machine with each added qubit.
}
\label{scaling2}
\end{figure}

\begin{figure}[ht]
\begin{center}
\includegraphics[width=0.85\hsize]{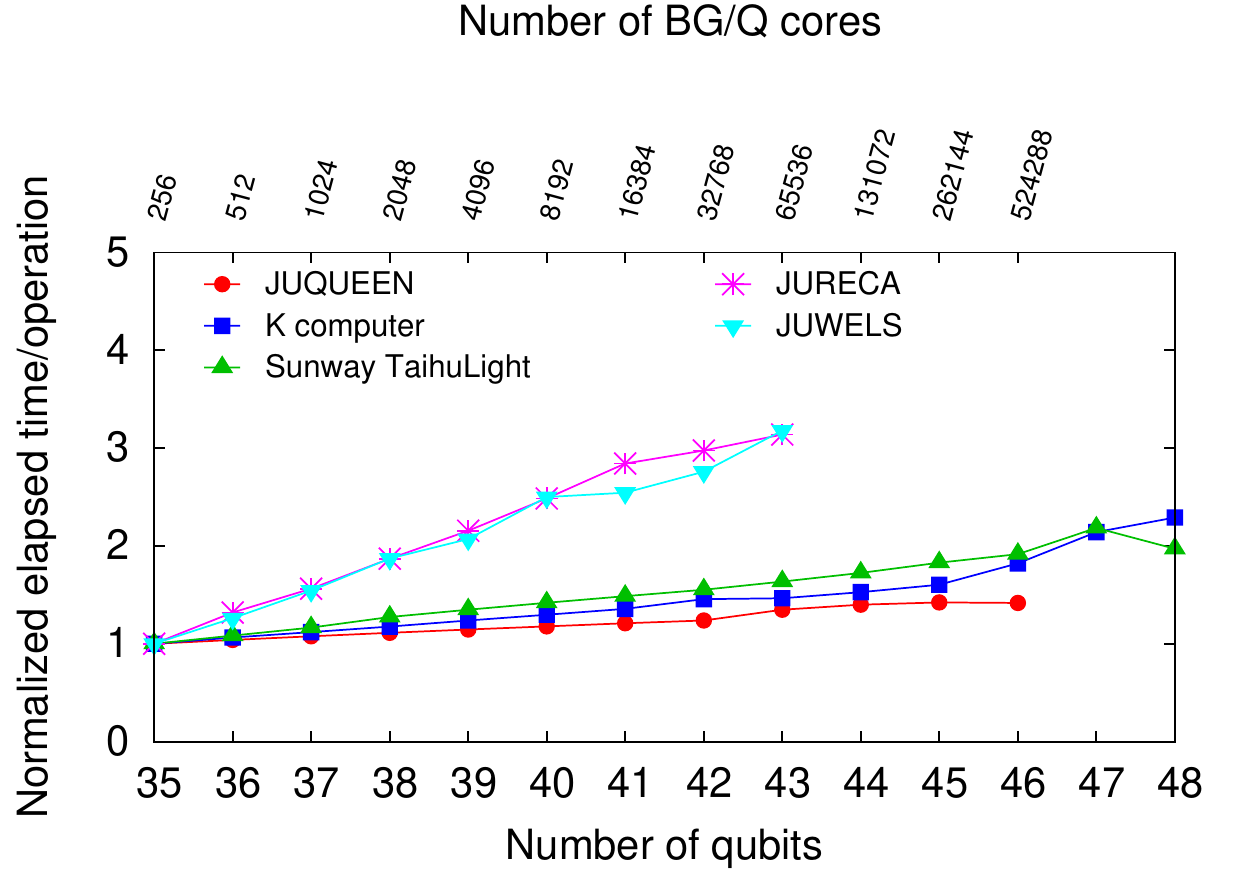}
\end{center}
\caption{The elapsed time per gate operation
(normalized by the values 2.7 s (JUQUEEN), 3.8 s (K) , 19.9 s (Sunway TaihuLight), 2.4 s (JURECA), and 2.2 s (JUWELS)
to run the 35 qubits circuit)
as a function of the number of qubits, obtained by \AQC\ executing a Hadamard operation on qubit 0 and the sequence
(CNOT 0 1), (CNOT 1 2), ..., (CNOT N-2, N-1).
This weak scaling plot shows that \AQC\ beats the exponential
scaling of the computational work by doubling the size of the machine with each added qubit.
}
\label{scaling3}
\end{figure}

Table~\ref{CNOTtab} and Figs.~\ref{scaling2} and \ref{scaling3} summarize the
results obtained by executing the sequence of gate operations
(H 0), (CNOT 0 1), (CNOT 1 2), ..., (CNOT N-2, N-1). 
The result of this quantum circuit is to put the quantum computer
in the maximally entangled state $(|0\ldots0\rangle+|1\ldots1\rangle)/\sqrt{2}$.
Also for these circuits, by doubling the size of the machine with each added qubit, \QC\ beats
the exponential scaling of the computational work with the number of qubits,
the salient feature of a gate-based universal quantum computer.

Performing single- and two-qubit gates requires only a little amount of computation per two or four basis states, respectively.
Some gates, such as CNOT and X, only perform a permutation of the elements of the state.
In other words, the arithmetic intensity of these operations is very low and the performance is mainly limited by the memory bandwidth.
This explains that the weak scaling behavior of the circuit with the many CNOT's
is slightly worse than that of the circuit involving Hadamard gates only.
One strategy to overcome this limitation is to increase the arithmetic intensity
by combining single- and two-qubit gates to multi-qubit (say 5-qubit) gates.
As \QC\ is also a test bed for the simulator SPI12MPI for spin-1/2 models,
we refrained from implementing this rather specialized strategy in the present version of \QC.
While the weak scaling behavior on JURECA and JUWELS is rather good by itself,
it is not as good as the ones on the other supercomputers used.
This suggests that there may be some limitations in the bandwidth to the memory and network on JURECA and JUWELS.

The 2-byte encoding/decoding used by \AQC\ to reduce the amount of required memory comes at the cost of larger computation time,
affecting the ratio between computation and communications.
This extra time depends on the type of quantum gate and ranges from almost zero (e.g. CNOT gate) to a factor of 3--4 (e.g Hadamard gate).
As a result, comparing elapsed times of \AQC\ and \EQC\ only makes sense if we execute the same quantum circuit and even then,
because of the factor-of-eight difference in memory usage, interpreting the differences in these
elapsed times is not straightforward.

\begin{table*}[ht]
\caption{Results of summing two 19-bit integers ($210018$ and $314269$)
using a quantum adder circuit~\cite{DRAP00,RAED07x}.
The BIT ASSIGNMENT instruction is used to interchange qubits (0--18) and (19--37)
such that the sum of the integers (all 19 bits equal to one) is returned in qubits (0--18).
This operation also reduces the amount of MPI communication.
Recall that \AQC\ uses a factor of 8 less memory than \EQC\ and,
as some of the numbers in the left three columns show,
returns results that deviate slightly from the numerically exact results produced by \EQC.
Calculations were performed on JUQUEEN using 8192 cores and 8192 MPI processes
and took 1446 seconds for \AQC\ and 388 seconds for \EQC\ to complete.
}
\begin{center}
\begin{tabular}{c|ccc|ccc}
\hline
     &\multicolumn{3}{c|}{\AQC}&\multicolumn{3}{c}{\EQC}\\
qubit&$\langle Qx(i) \rangle$&$\langle Qy(i) \rangle$&$\langle Qz(i) \rangle$&
$\langle Qx(i) \rangle$&$\langle Qy(i) \rangle$&$\langle Qz(i) \rangle$ \\
\hline
    0 & 0.500 & 0.500 & 0.999 &0.500 & 0.500 & 1.000 \\
    1 & 0.500 & 0.504 & 0.999 &0.500 & 0.500 & 1.000 \\
    2 & 0.504 & 0.502 & 0.999 &0.500 & 0.500 & 1.000 \\
    3 & 0.500 & 0.500 & 0.999 &0.500 & 0.500 & 1.000 \\
    4 & 0.504 & 0.501 & 0.999 &0.500 & 0.500 & 1.000 \\
    5 & 0.504 & 0.499 & 1.000 &0.500 & 0.500 & 1.000 \\
    6 & 0.500 & 0.500 & 0.999 &0.500 & 0.500 & 1.000 \\
    7 & 0.500 & 0.500 & 1.000 &0.500 & 0.500 & 1.000 \\
    8 & 0.500 & 0.500 & 1.000 &0.500 & 0.500 & 1.000 \\
    9 & 0.500 & 0.500 & 1.000 &0.500 & 0.500 & 1.000 \\
   10 & 0.500 & 0.500 & 1.000 &0.500 & 0.500 & 1.000 \\
   11 & 0.500 & 0.500 & 1.000 &0.500 & 0.500 & 1.000 \\
   12 & 0.500 & 0.500 & 1.000 &0.500 & 0.500 & 1.000 \\
   13 & 0.500 & 0.500 & 1.000 &0.500 & 0.500 & 1.000 \\
   14 & 0.500 & 0.500 & 1.000 &0.500 & 0.500 & 1.000 \\
   15 & 0.500 & 0.500 & 1.000 &0.500 & 0.500 & 1.000 \\
   16 & 0.489 & 0.497 & 1.000 &0.500 & 0.500 & 1.000 \\
   17 & 0.500 & 0.500 & 1.000 &0.500 & 0.500 & 1.000 \\
   18 & 0.500 & 0.500 & 1.000 &0.500 & 0.500 & 1.000 \\
   19 & 0.500 & 0.500 & 0.000 &0.500 & 0.500 & 0.000 \\
   20 & 0.500 & 0.500 & 1.000 &0.500 & 0.500 & 1.000 \\
   21 & 0.500 & 0.500 & 1.000 &0.500 & 0.500 & 1.000 \\
   22 & 0.500 & 0.500 & 0.000 &0.500 & 0.500 & 0.000 \\
   23 & 0.500 & 0.500 & 0.000 &0.500 & 0.500 & 0.000 \\
   24 & 0.500 & 0.500 & 1.000 &0.500 & 0.500 & 1.000 \\
   25 & 0.500 & 0.500 & 1.000 &0.500 & 0.500 & 1.000 \\
   26 & 0.500 & 0.500 & 0.000 &0.500 & 0.500 & 0.000 \\
   27 & 0.500 & 0.500 & 1.000 &0.500 & 0.500 & 1.000 \\
   28 & 0.500 & 0.500 & 0.000 &0.500 & 0.500 & 0.000 \\
   29 & 0.500 & 0.500 & 0.000 &0.500 & 0.500 & 0.000 \\
   30 & 0.500 & 0.500 & 0.000 &0.500 & 0.500 & 0.000 \\
   31 & 0.500 & 0.500 & 1.000 &0.500 & 0.500 & 1.000 \\
   32 & 0.500 & 0.500 & 1.000 &0.500 & 0.500 & 1.000 \\
   33 & 0.500 & 0.500 & 0.000 &0.500 & 0.500 & 0.000 \\
   34 & 0.500 & 0.500 & 0.000 &0.500 & 0.500 & 0.000 \\
   35 & 0.500 & 0.500 & 0.000 &0.500 & 0.500 & 0.000 \\
   36 & 0.500 & 0.500 & 1.000 &0.500 & 0.500 & 1.000 \\
   37 & 0.500 & 0.500 & 0.000 &0.500 & 0.500 & 0.000 \\
\hline
\end{tabular}
\label{ADDtab}
\end{center}
\end{table*}

\subsection{Adder circuit}

The quantum circuit that performs the addition (modulo $2^K$) of $M$ integers~\cite{DRAP00,RAED07x,MICH17b},
each represented by $K$ qubits, provides a simple, scalable, and easy-to-verify algorithm to validate
universal quantum computer simulators~\cite{MICH17b}.
It involves a quantum Fourier transform~\cite{NIEL10} and primarily performs controlled-phase gates.
We have constructed and executed quantum circuits that add up to five 9-bit integers
and have run a sample of these circuits on the K computer and BG/Q.

Table~\ref{ADDtab} shows some representative results of a quantum circuit that adds two 19-bit integers.
These results have been obtained by JUQUEEN using 8192 cores and 8192 MPI processes
and took 1446 seconds for \AQC\ and 388 seconds for \EQC\ to complete.
In this example, the values of the integers ($210018$ and $314269$)
are chosen such that their sum ($2^{19}-1$) corresponds to a binary
number with 19 bits equal to one, which makes it very easy to verify the correctness of the result.
If the quantum circuit works properly, the expectation values of the corresponding qubits should be equal to one.
Clearly, Table~\ref{ADDtab} confirms that \EQC\ works properly and also shows that
the results of \AQC\ are close but, in contrast to what the examples presented earlier might suggest,
not always equal to the numerically exact results.

\subsection{Shor's algorithm on a 48 qubit quantum computer}\label{SHOR}

For a detailed description of this algorithm, see Ref.~\onlinecite{NIEL10,SHOR99}.
Briefly, Shor's algorithm finds the prime factors $p$ and $q$ of a composite
integer $G=p\times q$ by determining the period of the function $f(x)=y^x\mod G$ for $x=0,1,\ldots $ Here, $1< y< G$
should be coprime (greatest common divisor of $y$ and $G$ is 1) to $G$.
If, by accident, $y$ and $G$ were not coprimes then $y=p$ or $y=q$ and there is no need to continue with the algorithm.
Let $r$ denote the period of $f(x)$, that is $f(x)=f(x+r)$.
If the chosen value of $y$ yields an odd period $r$, we repeat the algorithm with another choice for $y$,
until we find an $r$ that is even. Once we have found an even period $r$, we compute $y^{r/2}\mod G$.
If $y^{r/2}\neq \pm 1\mod G$, then we find the factors of $G$ by calculating the greatest common divisors of
$y^{r/2}\pm 1$ and $G$.

\begin{figure}[ht]
\begin{center}
\includegraphics[width=0.85\hsize]{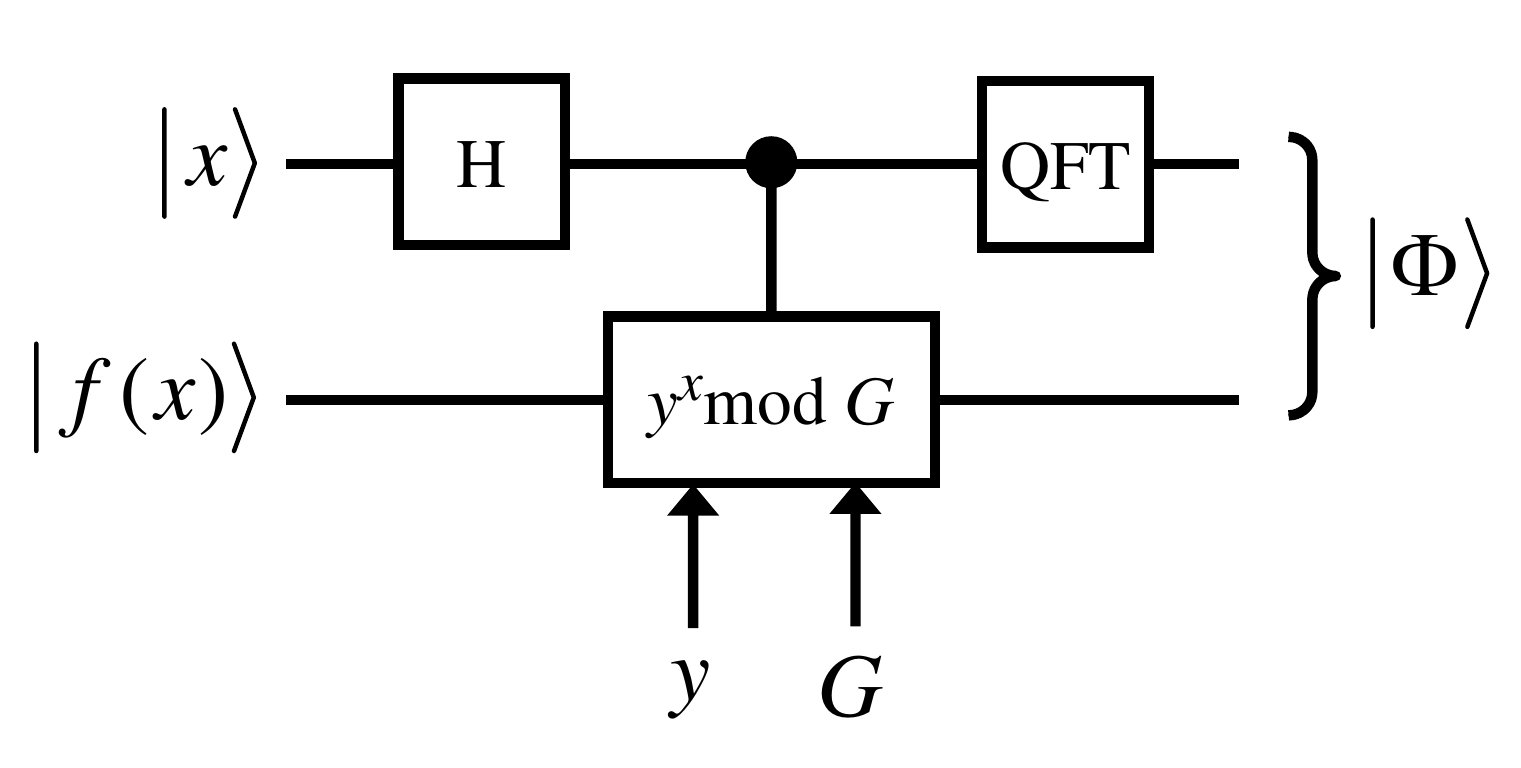}
\end{center}
\caption{Schematic diagram of Shor's algorithm.
}
\label{Shor}
\end{figure}

The schematic diagram of Shor's algorithm is shown in Fig.~\ref{Shor}. The quantum computer has $N$ qubits.
There are two qubit registers: An $x$-register with $X$ qubits
to hold the values of $x$- and a $f$-register with $F=N-X$ qubits to hold the values of $f(x)=y^x\mod G$.

What is the largest number Shor's original algorithm can factorize on a quantum computer with $N$ qubits?
The number of qubits to represent $y^x\mod G$ is $F=\lceil\log_2 G\rceil$.
For Shor's algorithm to work properly, that is to find the correct period $r$ of $f(x)$,
the number of qubits $X$ in the $x$-register should satisfy $G^2\leq 2^{X}< 2G^2$~\cite{SHOR99}.
Omitting numbers $G$ that can be written as a power of two (which are trivial to factorize),
the minimum number of qubits of the $x$-register is $X=\lceil\log_2 G^2\rceil$, so $N=X+F$ is either $3F$ or $3F-1$.
It follows that the maximum number of qubits that can be reserved for the $f$-register is given by $F=\lfloor (N+1)/3\rfloor$,
which determines the largest value of $G$.
For example, on a 45-, or 46-qubit quantum computer $G=32765$ is the largest integer
composed of two primes that can be factorized by Shor's algorithm.

The SHORBOX instruction of \QC\ takes $G$ and $y$ as input,
performs the Hadamard operations on all qubits of the $x$-register
and also computes $f(x)=y^x\mod G$ conditional on the qubits in the $x$-register and stores
the result in the $f$-register.
Application of the quantum Fourier transform on the $x$-register and sampling the state in the $x$-register
produces numbers of basis states which can then be used to determine the period $r$ and
the factors $p$ and $q$~\cite{NIEL10,SHOR99}.
The task of \QC\ is to execute SHORBOX and perform the quantum Fourier transform.

In Table~\ref{Shortab1}, we present the results for the case $G=1007$ and $y=529$, using
30 qubits, as obtained by running the \AQC\ on a Lenovo W520 notebook.
For comparison, we show the expectation values of the three components of the qubits
as given by \QC\ together with the exact results calculated from the exact closed-form expression~\cite{RAED07x}.
The results of \AQC\ agree very well with the exact ones.

\begin{table*}[ht]
\caption{Representative results of running Shor's algorithm with \AQC\
for a 30 qubit QC, $G=1007=19\times53$ and $y=529$, yielding a period $r=18$.
The three rightmost columns give the exact results, obtained from the closed-form expression~\cite{RAED07x}
of the expectation values of the individual qubits.
The expectation values produced by \EQC\ are numerically exact and are therefore not shown.
The number of qubits in the $x$-register is 20.
The calculation used all 8 cores of a Lenovo W520 notebook (Windows 10) and took 348 seconds (elapsed time) to complete.
}
\begin{center}
\begin{tabular}{c|ccc|ccc}
\hline
     &\multicolumn{3}{c|}{\AQC}&\multicolumn{3}{c}{Exact}\\
qubit&$\langle Qx(i) \rangle$&$\langle Qy(i) \rangle$&$\langle Qz(i) \rangle$&
$\langle Qx(i) \rangle$&$\langle Qy(i) \rangle$&$\langle Qz(i) \rangle$ \\
\hline
    0 & 0.504 & 0.495 & 0.500 & 0.500 & 0.500  &  0.500  \\
    1 & 0.502 & 0.497 & 0.500 & 0.500 & 0.500  &  0.500  \\
    2 & 0.502 & 0.499 & 0.500 & 0.500 & 0.500  &  0.500  \\
    3 & 0.502 & 0.499 & 0.445 & 0.500 & 0.500  &  0.445  \\
    4 & 0.504 & 0.500 & 0.445 & 0.500 & 0.500  &  0.445  \\
    5 & 0.506 & 0.500 & 0.445 & 0.500 & 0.500  &  0.445  \\
    6 & 0.501 & 0.500 & 0.444 & 0.500 & 0.500  &  0.444  \\
    7 & 0.500 & 0.500 & 0.444 & 0.500 & 0.500  &  0.444  \\
    8 & 0.500 & 0.500 & 0.444 & 0.500 & 0.500  &  0.444  \\
    9 & 0.500 & 0.500 & 0.444 & 0.500 & 0.500  &  0.444  \\
   10 & 0.500 & 0.500 & 0.444 & 0.500 & 0.500  &  0.444  \\
   11 & 0.500 & 0.500 & 0.444 & 0.500 & 0.500  &  0.444  \\
   12 & 0.500 & 0.500 & 0.444 & 0.500 & 0.500  &  0.444  \\
   13 & 0.501 & 0.500 & 0.444 & 0.500 & 0.500  &  0.444  \\
   14 & 0.501 & 0.500 & 0.444 & 0.500 & 0.500  &  0.444  \\
   15 & 0.500 & 0.500 & 0.444 & 0.500 & 0.500  &  0.444  \\
   16 & 0.501 & 0.500 & 0.444 & 0.500 & 0.500  &  0.444  \\
   17 & 0.500 & 0.500 & 0.444 & 0.500 & 0.500  &  0.444  \\
   18 & 0.501 & 0.499 & 0.444 & 0.500 & 0.500  &  0.444  \\
   19 & 0.500 & 0.500 & 0.500 & 0.500 & 0.500  &  0.500  \\
\hline
\end{tabular}
\label{Shortab1}
\end{center}
\end{table*}

In Table~\ref{Shortab2}, we present the results for the case $G=32399$ and $y=4295$, using
45 qubits, as obtained by running the \AQC\ on JUQUEEN.
For comparison, we show the expectation values of the three components of the qubits
as given by \QC\ together with the exact results calculated from the exact closed-form expression~\cite{RAED07x}.
The results of \AQC\ agree very well with the exact ones.
We conjecture that a comparable accuracy of about 3 digits on the expectation values of single qubits
($\langle Qz(i) \rangle$)
is beyond the reach of present~\cite{MICH17b} and future hardware realizations of gate-based quantum computers.

On a 48- or 49-qubit quantum computer, the largest composite integer $G=p\times q$ that can be factorized with
Shor's original algorithm is $G=65531=19\times3449$.
We employed \AQC\ to run Shor's algorithm on a 48 qubit universal quantum computer (simulator)
for $G=64507$ and $G=65531$, requiring 32 qubits for the $x$-register and 16 qubits for the $f$-register.
On the Sunway TaihuLight, we made a run with $y=21587$ and, after about 347 minutes of elapsed time,
obtained a result that shows a period $r=2$, which according to Shor's algorithm, yields
the factorization $64507=251\times257$.
On the K computer, a run with $y=34888$ yielded, after 299 minutes of elapsed time,
a result with period $r=4$. According to Shor's algorithm this implies that $64507=251\times257$,
in concert with the result obtained on the Sunway TaihuLight.
Running Shor's algorithm with $G=65531$ and $y=1122$ on the K computer returned
after 300 minutes of elapsed time, a result with period $r=4$,
in agreement with $G=65531=19\times3449$.

\begin{table*}[ht]
\caption{Representative results of running Shor's algorithm with \AQC\
for a 45 qubit QC, $G=32399=179\times181$ and $y=4295$, yielding a period $r=6$.
The three rightmost columns give the exact results, obtained from the closed-form expression~\cite{RAED07x}
of the expectation values of the individual qubits.
The number of qubits on the $x$-register is 30.
The calculation was performed on JUQUEEN, using 262144 cores, and took 5669 seconds of elapsed time to complete.
}
\begin{center}
\begin{tabular}{c|ccc|ccc}
\hline
     &\multicolumn{3}{c|}{\AQC}&\multicolumn{3}{c}{Exact}\\
qubit&$\langle Qx(i) \rangle$&$\langle Qy(i) \rangle$&$\langle Qz(i) \rangle$&
$\langle Qx(i) \rangle$&$\langle Qy(i) \rangle$&$\langle Qz(i) \rangle$ \\
\hline
    0 & 0.499 & 0.480 & 0.502  & 0.500 & 0.500  &  0.500 \\
    1 & 0.579 & 0.486 & 0.375  & 0.500 & 0.500  &  0.375 \\
    2 & 0.494 & 0.494 & 0.345  & 0.500 & 0.500  &  0.344 \\
    3 & 0.518 & 0.498 & 0.336  & 0.500 & 0.500  &  0.336 \\
    4 & 0.496 & 0.498 & 0.335  & 0.500 & 0.500  &  0.334 \\
    5 & 0.503 & 0.500 & 0.333  & 0.500 & 0.500  &  0.333 \\
    6 & 0.499 & 0.499 & 0.335  & 0.500 & 0.500  &  0.333 \\
    7 & 0.502 & 0.500 & 0.333  & 0.500 & 0.500  &  0.333 \\
    8 & 0.500 & 0.500 & 0.334  & 0.500 & 0.500  &  0.333 \\
    9 & 0.500 & 0.500 & 0.333  & 0.500 & 0.500  &  0.333 \\
   10 & 0.500 & 0.500 & 0.334  & 0.500 & 0.500  &  0.333 \\
   11 & 0.501 & 0.500 & 0.333  & 0.500 & 0.500  &  0.333 \\
   12 & 0.500 & 0.500 & 0.335  & 0.500 & 0.500  &  0.333 \\
   13 & 0.499 & 0.500 & 0.333  & 0.500 & 0.500  &  0.333 \\
   14 & 0.501 & 0.500 & 0.335  & 0.500 & 0.500  &  0.333 \\
   15 & 0.499 & 0.500 & 0.333  & 0.500 & 0.500  &  0.333 \\
   16 & 0.501 & 0.500 & 0.335  & 0.500 & 0.500  &  0.333 \\
   17 & 0.499 & 0.500 & 0.333  & 0.500 & 0.500  &  0.333 \\
   18 & 0.500 & 0.500 & 0.335  & 0.500 & 0.500  &  0.333 \\
   19 & 0.499 & 0.500 & 0.333  & 0.500 & 0.500  &  0.333 \\
   20 & 0.501 & 0.500 & 0.335  & 0.500 & 0.500  &  0.333 \\
   21 & 0.499 & 0.500 & 0.333  & 0.500 & 0.500  &  0.333 \\
   22 & 0.501 & 0.500 & 0.335  & 0.500 & 0.500  &  0.333 \\
   23 & 0.499 & 0.500 & 0.333  & 0.500 & 0.500  &  0.333 \\
   24 & 0.500 & 0.500 & 0.335  & 0.500 & 0.500  &  0.333 \\
   25 & 0.499 & 0.500 & 0.332  & 0.500 & 0.500  &  0.333 \\
   26 & 0.501 & 0.500 & 0.335  & 0.500 & 0.500  &  0.333 \\
   27 & 0.500 & 0.500 & 0.332  & 0.500 & 0.500  &  0.333 \\
   28 & 0.500 & 0.500 & 0.335  & 0.500 & 0.500  &  0.333 \\
   29 & 0.500 & 0.500 & 0.500  & 0.500 & 0.500  &  0.500 \\
\hline
\end{tabular}
\label{Shortab2}
\end{center}
\end{table*}

\section{Conclusion}\label{section6}

The revised version of the massively parallel quantum computer simulator has been
used to run a variety of quantum circuits on the Sunway TaihuLight, on the K computer, on an IBM Blue Gene/Q,
and on Intel Xeon based clusters.
Close-to-linear weak scaling of the elapsed time as a function of the number of qubits was observed
on all computers used.
This implies that the combination of software, many cores, and a fast communication network
beats the exponential increase in memory and CPU time that is the characteristic of
simulating quantum systems on a digital computer.

Two techniques for alleviating the memory problem have been discussed.
The first employs an adaptive coding scheme to represent the
quantum state in terms of 2-byte instead of 16-byte numbers. Benchmarks
including Shor's algorithm, adders, quantum Fourier transforms, Hadamard and CNOT operations
show that the factor-of-eight reduction in memory
has no significant impact on the accuracy of the outcomes.
This version can simulate a 32-qubit universal quantum computer on a notebook with 16 GB of memory.

The second technique resorts to a well-known method of Quantum Monte Carlo simulations to
express two-qubit gates in terms of single-qubit gates and auxiliary variables.
The worst case run time and memory usage of this algorithm was shown to be ${\cal O}(N M 2^P)$
and ${\cal O}(N+M)$, respectively, where $N$ is the number of qubits,
$P$ is the number of two-qubit gates and $M$ is the number of output amplitudes desired.
Although the reduction in memory can be huge if $M\ll N$, the technique is of limited practical
use unless one knows how to choose $M$ basis states of interest.

Through the specification of the sequence of quantum gates, the input to \QC\
can easily be tailored to put a heavy burden on the communication network, memory, processor or any combination of them.
Therefore, the simulator described in this paper may be a useful addition to the suite of benchmarks for new high-performance computers.
As mentioned in the introduction, the current version was designed to be portable over a wide range of computing platforms.
However, the new generation of high-performance computers rely on accelerators or GPUs to deliver higher performance.
For instance, the Sunway TaihuLight requires machine-specific programming to make use of the accelerator hardware.
Adapting the code to make efficient use of GPUs or other kinds of accelerators is a challenging project that we leave for future work.

\section*{Acknowledgements}
The authors thank Koen De Raedt for his help in improving the simulation code.
The authors acknowledge the computing time granted by the JARA-HPC Vergabegremium and provided on
the JARA-HPC Partition part of the supercomputer JUQUEEN~\cite{JUQUEEN} at the Forschungszentrum J\"ulich.
D.W. is supported by the Initiative and Networking Fund of the Helmholtz Association through the Strategic Future
Field of Research project ``Scalable solid state quantum computing (ZT-0013)''.
Part of the simulations reported in this paper were carried out on the K computer at RIKEN.
S. Yuan gratefully acknowledges financial support from the Thousand Young Talent Plan
and computational resources provided by National Supercomputing Center in Wuxi (China).

\appendix
\section{Instruction set}\label{APPb}

This appendix gives a detailed specification of each of the gate operations that are implemented in \QC.
\QC\ will become accessible through a J\"ulich cloud service in 2019.
A dockerized version of the four executables (\EQC\ and \AQC, MPI/OpenMP and MPI only) is available on request.

\medskip
\setlist[description]{leftmargin=\parindent,labelindent=\parindent}

\noindent{\bf I gate}
\begin{description}
\item[Description] the I gate performs an identity operation on qubit $n$.
\item[Syntax] I $n$
\item[Argument] $n$ is in the range $0,\dots,N-1$ where $N$ is the number of qubits.
\item[Operation] $I=\left(\begin{array}{rr} 1 & 0\\ 0 & 1 \end{array}\right)$
\item[Graphical symbol]\hskip 10pt $\Qcircuit @C=1.3em @R=.4em {&\gate{I}&\qw\\}$
\item[Note] The I gate is implemented as a ``no operation'' and is provided for compatibility with some other
assembler-like language only.

\end{description}
\medskip

\noindent{\bf H gate}
\begin{description}
\item[Description] the H gate performs a Hadamard operation on qubit $n$.
\item[Syntax] H $n$
\item[Argument] $n$ is in the range $0,\dots,N-1$ where $N$ is the number of qubits.
\item[Operation] $H=\frac{1}{\sqrt{2}}\left(\begin{array}{rr} 1 & 1\\ 1 & -1 \end{array}\right)$
\item[Graphical symbol]\hskip 10pt $\Qcircuit @C=1.3em @R=.4em {&\gate{H}&\qw\\}$
\end{description}
\medskip

\noindent{\bf X gate}
\begin{description}
\item[Description] the X gate performs a bit flip operation on qubit $n$.
\item[Syntax] X $n$
\item[Argument] $n$ is in the range $0,\dots,N-1$ where $N$ is the number of qubits.
\item[Operation] $X=\left(\begin{array}{rr} 0 & 1\\ 1 & 0 \end{array}\right)$
\item[Graphical symbol]\hskip 10pt $\Qcircuit @C=1.3em @R=.4em {&\gate{X}&\qw\\}$
\end{description}
\medskip

\noindent{\bf Y gate}
\begin{description}
\item[Description] the Y gate performs a bit and phase flip operation on qubit $n$.
\item[Syntax] Y $n$
\item[Argument] $n$ is in the range $0,\dots,N-1$ where $N$ is the number of qubits.
\item[Operation] $Y=\left(\begin{array}{rr} 0 & -i\\ i & 0 \end{array}\right)$
\item[Graphical symbol]\hskip 10pt $\Qcircuit @C=1.3em @R=.4em {&\gate{Y}&\qw\\}$
\end{description}
\medskip

\noindent{\bf Z gate}
\begin{description}
\item[Description] the Z gate performs a phase flip operation on qubit $n$.
\item[Syntax] Z $n$
\item[Argument] $n$ is in the range $0,\dots,N-1$ where $N$ is the number of qubits.
\item[Operation] $Z=\left(\begin{array}{rr} 1 & 0\\ 0 & -1 \end{array}\right)$
\item[Graphical symbol]\hskip 10pt $\Qcircuit @C=1.3em @R=.4em {&\gate{Z}&\qw\\}$
\end{description}
\medskip

\noindent{\bf S gate}
\begin{description}
\item[Description] the S gate rotates qubit $n$ about the $z$-axis by $\pi/4$.
\item[Syntax] S $n$
\item[Argument] $n$ is in the range $0,\dots,N-1$ where $N$ is the number of qubits.
\item[Operation] $S=\left(\begin{array}{rr} 1 & 0\\ 0 & i \end{array}\right)$
\item[Graphical symbol]\hskip 10pt $\Qcircuit @C=1.3em @R=.4em {&\gate{S}&\qw\\}$
\end{description}
\medskip

\noindent{\bf S$^\dagger$ gate}
\begin{description}
\item[Description] the S$^\dagger$ gate rotates qubit $n$ about the $z$-axis by $-\pi/4$
\item[Syntax] S+ $n$
\item[Argument] $n$ is in the range $0,\dots,N-1$ where $N$ is the number of qubits.
\item[Operation] $S^\dagger=\left(\begin{array}{rr} 1 & 0\\ 0 & -i \end{array}\right)$
\item[Graphical symbol]\hskip 10pt $\Qcircuit @C=1.3em @R=.4em {&\gate{S^\dagger}&\qw\\}$
\end{description}
\medskip

\noindent{\bf T gate}
\begin{description}
\item[Description] the T gate rotates qubit $n$ about the $z$-axis by $\pi/8$
\item[Syntax] T $n$
\item[Argument] $n$ is in the range $0,\dots,N-1$ where $N$ is the number of qubits.
\item[Operation] $T=\left(\begin{array}{cc} 1 & 0\\ 0 & (1+i)/\sqrt{2} \end{array}\right)$
\item[Graphical symbol]\hskip 10pt $\Qcircuit @C=1.3em @R=.4em {&\gate{T}&\qw\\}$
\end{description}
\medskip

\noindent{\bf T$^\dagger$ gate}
\begin{description}
\item[Description] the T$^\dagger$ gate rotates qubit $n$ about the $z$-axis by $-\pi/8$
\item[Syntax] T+ $n$
\item[Argument] $n$ is in the range $0,\dots,N-1$ where $N$ is the number of qubits.
\item[Operation] $T^\dagger=\left(\begin{array}{cc} 1 & 0\\ 0 & (1-i)/\sqrt{2}\end{array}\right)$
\item[Graphical symbol]\hskip 10pt $\Qcircuit @C=1.3em @R=.4em {&\gate{T^\dagger}&\qw\\}$
\end{description}
\medskip

\noindent{\bf U1 gate}
\begin{description}
\item[Description] the U1 gate performs a U1($\lambda$) operation~\cite{CROS17} on qubit $n$.
\item[Syntax] U1 $n$ $\lambda$
\item[Arguments] $n$ is an integer in the range $0,\dots,N-1$ where $N$ is the number of qubits and $\lambda$ is a number (floating point or integer) that represents an angle expressed in radians.
\item[Operation] $U1(\lambda)=\left(\begin{array}{cc} 1 & 0\\ 0 & e^{i\lambda} \end{array}\right)$
\item[Graphical symbol]\hskip 10pt $\Qcircuit @C=1.3em @R=.4em {&\gate{U1(\lambda)}&\qw\\}$
\end{description}
\medskip

\noindent{\bf U2 gate}
\begin{description}
\item[Description] the U2 gate performs a U2($\phi,\lambda$) operation~\cite{CROS17} on qubit $n$.
\item[Syntax] U2 $n$ $\phi$ $\lambda$
\item[Arguments] $n$ is an integer in the range $0,\dots,N-1$ where $N$ is the number of qubits and $\phi$
and $\lambda$ are numbers (floating point or integer) that represent angles expressed in radians.
\item[Operation] $U2(\phi,\lambda)=\frac{1}{\sqrt{2}}
\left(\begin{array}{cc} 1 & -e^{i\lambda}\\ e^{i\phi} & e^{i(\phi+\lambda)} \end{array}\right)$
\item[Graphical symbol]\hskip 10pt $\Qcircuit @C=1.3em @R=.4em {&\gate{U2(\phi,\lambda)}&\qw\\}$
\end{description}
\medskip

\noindent{\bf U3 gate}
\begin{description}
\item[Description] the U3 gate performs a U3($\theta,\phi,\lambda$) operation~\cite{CROS17} on qubit $n$.
\item[Syntax] U3 $n$ $\theta$ $\phi$ $\lambda$
\item[Arguments] $n$ is an integer in the range $0,\dots,N-1$ where $N$ is the number of qubits and $\theta$, $\phi$ and $\lambda$ are numbers (floating point or integer) that represent angles expressed in radians.
\item[Operation] $U3(\theta,\phi,\lambda)=
\left(\begin{array}{cc} \cos(\theta/2) &  -e^{i\lambda}\sin(\theta/2)\\ e^{i\phi}\sin(\theta/2)& e^{i(\phi+\lambda)}\cos(\theta/2) \end{array}\right)$
\item[Graphical symbol]\hskip 10pt $\Qcircuit @C=1.3em @R=.4em {&\gate{U3(\theta,\phi,\lambda)}&\qw\\}$
\end{description}
\medskip

\noindent{\bf +X gate}
\begin{description}
\item[Description] the +X gate rotates qubit $n$ by $-\pi/2$ about the $x$-axis.
\item[Syntax] +X $n$
\item[Argument] $n$ is in the range $0,\dots,N-1$ where $N$ is the number of qubits.
\item[Operation] $+X=\frac{1}{\sqrt{2}}\left(\begin{array}{rr} 1 & i\\ i & 1 \end{array}\right)$
\item[Graphical symbol]\hskip 10pt $\Qcircuit @C=1.3em @R=.4em {&\gate{+X}&\qw\\}$
\end{description}
\medskip

\noindent{\bf -X gate}
\begin{description}
\item[Description] the -X gate rotates qubit $n$ by $+\pi/2$ about the $x$-axis.
\item[Syntax] -X $n$
\item[Argument] $n$ is in the range $0,\dots,N-1$ where $N$ is the number of qubits.
\item[Operation] $-X=\frac{1}{\sqrt{2}}\left(\begin{array}{rr} 1 & -i\\ -i & 1 \end{array}\right)$
\item[Graphical symbol]\hskip 10pt $\Qcircuit @C=1.3em @R=.4em {&\gate{-X}&\qw\\}$
\end{description}
\medskip

\noindent{\bf +Y gate}
\begin{description}
\item[Description] the +Y gate rotates qubit $n$ by $-\pi/2$ about the $y$-axis.
\item[Syntax] +Y $n$
\item[Argument] $n$ is in the range $0,\dots,N-1$ where $N$ is the number of qubits.
\item[Operation] $+Y=\frac{1}{\sqrt{2}}\left(\begin{array}{rr} 1 & 1\\ -1 & 1 \end{array}\right)$
\item[Graphical symbol]\hskip 10pt $\Qcircuit @C=1.3em @R=.4em {&\gate{+Y}&\qw\\}$
\end{description}
\medskip

\noindent{\bf -Y gate}
\begin{description}
\item[Description] the -Y gate rotates qubit $n$ by $+\pi/2$ about the $y$-axis.
\item[Syntax] -Y $n$
\item[Argument] $n$ is in the range $0,\dots,N-1$ where $N$ is the number of qubits.
\item[Operation] $-Y=\frac{1}{\sqrt{2}}\left(\begin{array}{rr} 1 & -1\\ 1 & 1 \end{array}\right)$
\item[Graphical symbol]\hskip 10pt $\Qcircuit @C=1.3em @R=.4em {&\gate{-Y}&\qw\\}$
\end{description}
\medskip

\noindent{\bf R(k) gate}
\begin{description}
\item[Description] the R(k) gate changes the phase of qubit $n$ by an angle $2\pi/2^k$.
\item[Syntax] R $n$ $k$
\item[Arguments] $n$ is in the range $0,\dots,N-1$ where $N$ is the number of qubits and $k$ is a non-negative integer.
\item[Operation] $R(k)=\left(\begin{array}{cc} 1 & 0\\ 0 & e^{2\pi i/2^k} \end{array}\right)$
\item[Graphical symbol]\hskip 10pt $\Qcircuit @C=1.3em @R=.4em {&\gate{R(k)}&\qw\\}$
\end{description}
\medskip

\noindent{\bf R$^\dagger$(k) gate}
\begin{description}
\item[Description] the inverse R(k) gate changes the phase of qubit $n$ by an angle $-2\pi/2^k$.
\item[Syntax] R $n$ $-k$
\item[Arguments] $n$ is in the range $0,\dots,N-1$ where $N$ is the number of qubits and $k$ is a non-negative integer.
\item[Operation] $R^\dagger(k)=\left(\begin{array}{cc} 1 & 0\\ 0 & e^{-2\pi i/2^k} \end{array}\right)$
\item[Graphical symbol]\hskip 10pt $\Qcircuit @C=1.3em @R=.4em {&\gate{R^\dagger(k)}&\qw\\}$
\end{description}
\medskip


\noindent{\bf CNOT gate}
\begin{description}
\item[Description] the controlled-NOT gate flips the target qubit if the control qubit is 1.
\item[Syntax] CNOT $control$ $target$
\item[Arguments] $control\not=target$ are integers in the range $0,\dots,N-1$ where $N$ is the number of qubits.
\item[Operation] $\mathrm{CNOT}=\left(\begin{array}{cccc} 1 & 0 & 0 & 0\\ 0 & 1 & 0 & 0 \\ 0 & 0 & 0 & 1\\0 & 0 & 1 & 0\end{array}\right)$
\ ; in the computational basis $|control,target\rangle$.
\item[Graphical symbol]\hskip 10pt $\Qcircuit @C=1.3em @R=.4em {&\ctrl{1}&\qw\\&\targ&\qw}$
\end{description}
\medskip

\noindent{\bf U(k) gate}
\begin{description}
\item[Description] the controlled-phase gate shifts the phase of the target qubit by an angle $2\pi/2^k$ if the control qubit is 1.
\item[Syntax] U $control$ $target$ $k$
\item[Arguments] $control\not=target$ are integers in the range $0,\dots,N-1$ where $N$ is the number of qubits
and $k$ is a non-negative integer.
\item[Operation] $U(k)=\left(\begin{array}{cccc}
1 & 0 & 0 & 0\\
0 & 1 & 0 & 0\\
0 & 0 & 1 & 0\\
0 & 0 & 0 & e^{2\pi i/2^k}
\end{array}\right)$
\ ; in the computational basis $|control,target\rangle$.
\item[Graphical symbol]\hskip 10pt $\Qcircuit @C=1.3em @R=.4em {&\ctrl{1}&\qw\\&\gate{U(k)}&\qw}$
\end{description}
\medskip

\noindent{\bf U$^\dagger$(k) gate}
\begin{description}
\item[Description] the U$^\dagger$ gate shifts the phase of the target qubit by an angle $-2\pi/2^k$ if the control qubit is 1.
\item[Syntax] U $control$ $target$ $-k$
\item[Arguments] $control\not=target$ are integers in the range $0,\dots,N-1$ where $N$ is the number of qubits and $k$
is a non-negative integer.
\item[Operation] $U^\dagger(k)=\left(\begin{array}{cccc}
1 & 0 & 0 & 0\\
0 & 1 & 0 & 0\\
0 & 0 & 1 & 0\\
0 & 0 & 0 & e^{-2\pi i/2^k}
\end{array}\right)$
\ in the computational basis $|control,target\rangle$.
\item[Graphical symbol]\hskip 10pt $\Qcircuit @C=1.3em @R=.4em {&\ctrl{1}&\qw\\&\gate{U^\dagger(k)}&\qw}$
\end{description}
\medskip


\noindent{\bf Toffoli gate}
\begin{description}
\item[Description] the TOFFOLI gate flips the target qubit if both control qubits are 1.
\item[Syntax] TOFFOLI $control_1$ $control_2$ $target$
\item[Arguments] $control_1\not=control_2\not=target\not=control_1$ are integers in the range $0,\dots,N-1$  where $N$ is the number of qubits.
\item[Operation] $\mathrm{TOFFOLI}=\left(\begin{array}{cccccccc}
1 & 0 & 0 & 0& 0 & 0 & 0& 0 \\
0 & 1 & 0 & 0& 0 & 0 & 0& 0 \\
0 & 0 & 1 & 0& 0 & 0 & 0& 0 \\
0 & 0 & 0 & 1& 0 & 0 & 0& 0 \\
0 & 0 & 0 & 0& 1 & 0 & 0& 0 \\
0 & 0 & 0 & 0& 0 & 1 & 0& 0 \\
0 & 0 & 0 & 0& 0 & 0 & 0& 1 \\
0 & 0 & 0 & 0& 0 & 0 & 1& 0 \\
\end{array}\right)$\ in the computational basis $|control_1,control_2,target\rangle$.
\item[Graphical symbol]\hskip 10pt $\Qcircuit @C=1.3em @R=.4em {&\ctrl{1}&\qw\\ &\ctrl{1}&\qw\\ &\targ&\qw}$
\end{description}
\medskip


\noindent{\bf BEGIN MEASUREMENT}
\begin{description}
\item[Description] BEGIN MEASUREMENT computes and prints out the expectation values of all $N$ qubits.
This operation does not change the state of the quantum computer.
\item[Syntax] BEGIN MEASUREMENT
\item[Arguments] None
\item[Operation] In terms of their representation in terms of Pauli matrices,
\QC\ computes
$\langle Qx(n)\rangle =\langle \Psi|(1-\sigma^x_n)|\Psi \rangle/2$,
$\langle Qy(n)\rangle =\langle \Psi|(1-\sigma^y_n)|\Psi \rangle/2$, and
$\langle Qz(n)\rangle =\langle \Psi|(1-\sigma^z_n)|\Psi \rangle/2$
for $n=0,\ldots,N-1$, where $|\Psi \rangle$ is the state of the quantum computer
at the time that BEGIN MEASUREMENT was issued.
\item[Graphical symbol]\hskip 10pt $\Qcircuit @C=1.3em @R=.4em {&\gate{\langle . \rangle}&\qw}$
\end{description}
\medskip

\noindent{\bf GENERATE EVENTS}
\begin{description}
\item[Description] GENERATE EVENTS computes the probabilities of each of the basis states.
It then uses random numbers to generate and print out the states according to these probabilities.
This operation destroys the state of the quantum computer. It will force \QC\ to exit.
\item[Syntax] GENERATE EVENTS $events$ $seed$
\item[Arguments] $events$ is a positive integer, determining the number of events that will be generated and
$seed$ is an integer that is used as the initial seed for the random number generator if $seed>0$.
If $seed\le0$, \QC\ uses as seed the value provided by the operating system.
\item[Operation] GENERATE EVENTS produces a list of $events$ states, all sampled from the probability distribution
computed from the current state of the quantum computer.
\item[Graphical symbol] none
\end{description}
\medskip

\noindent{\bf M}
\begin{description}
\item[Description] M performs a projective measurement on qubit $n$.
\item[Syntax] M $n$
\item[Arguments] $n$ is in the range $0,\dots,N-1$ where $N$ is the number of qubits.
\item[Operation] \QC\ first computes the probabilities $p_0$ and $p_1$
to observe qubit $n$ in the state $|0\rangle$ and $|1\rangle$, respectively,
Then \QC\ selects the measurement outcome 0 or 1 at random according
to these probabilities and projects the qubit onto the corresponding state.
\item[Graphical symbol]\hskip 10pt $\Qcircuit @C=1.3em @R=.4em {&\meter&\qw}$
\end{description}
\medskip


\noindent{\bf QUBITS}
\begin{description}
\item[Description] QUBITS specifies the number of qubits of the universal quantum computer.
\item[Syntax] QUBITS $N$
\item[Arguments] $N$ is an integer which must be larger than 1 and smaller than 64 (the actual number is
limited by the available memory).
\item[Note] QUBITS $N$ must be the first instruction.
\end{description}
\medskip

\noindent{\bf BIT ASSIGNMENT}
\begin{description}
\item[Description]
Applications that require MPI to run \QC\ on a distributed memory machine may benefit
from renumbering the qubits such that the amount of MPI communications is reduced.
The box below shows how this can be done without changing the original
quantum circuit, for a simulation involving 4 qubits.
\item[Syntax] BIT ASSIGNMENT  Permutation(0,1,\ldots,N-1), see Example 3.
\item[Arguments] A list of integers in the range $0,\ldots,N-1$ that is a permutation of the set $\{0,1,\ldots,N-1\}$.

\begin{center}
\vbox{
\smallskip
\smallskip
\framebox{
\parbox[ht]{0.85\hsize}{
\vskip 2pt
QUBITS         4

\noindent
BIT ASSIGNMENT  2 3 1 0
\vskip 2pt
}}}
\end{center}
\item[Note] This instruction should appear after QUBITS $N$ and before the first gate instruction.
\end{description}
\medskip

\noindent{\bf SHORBOX}
\begin{description}
\item[Description] SHORBOX initializes the $x$-register and the $f$-register
in Shor's algorithm~\cite{NIEL10} to the state of uniform superposition and $y^x \mod G$, respectively.
Here $G$ is the number to be factorized and $1<y<G$ is chosen to be coprime to $G$.
Subsequent application of the quantum Fourier transform to the $x$-register allows
for the determination of the period of the function $f(x)=y^x \mod G$
from which the factors of $G$ may be calculated~\cite{NIEL10,RAED07x}.
\item[Syntax] SHORBOX $n_x$ $G$ $y$
\item[Arguments] $n_x<N$ is the number of qubits reserved for the $x$-register,
and $G$ and $y$ are integers. See Ref.~\onlinecite{RAED07x} for details.
\item[Operation] $\mathrm{SHORBOX}=2^{-n_x/2}\sum_{x=0}^{2^{n_x-1}}|x\rangle |y^x \mod G\rangle$.
\item[Graphical symbol] example for $n_x=4$
\hskip 10pt \Qcircuit @C=1em @R=0em {
& \multigate{3}{\mathrm{Shor}} & \qw \\
& \ghost{\mathrm{Shor}} & \qw \\
& \ghost{\mathrm{Shor}} & \qw \\
& \ghost{\mathrm{Shor}} & \qw
}
\end{description}
\medskip

\noindent{\bf CLEAR}
\begin{description}
\item[Description] The CLEAR instruction projects the state of qubit $n$ to $|0\rangle$.
\item[Syntax] CLEAR $n$
\item[Arguments] $n$ is in the range $0,\dots,N-1$ where $N$ is the number of qubits.
\item[Operation] $\mathrm{CLEAR}=|0\rangle_n\langle 0|_n$
\item[Graphical symbol]\hskip 10pt $\Qcircuit @C=1.3em @R=.4em {&\gate{0}&\qw}$
\item[Note] This instruction fails if the projection results in to a state with amplitude zero.
\end{description}
\medskip

\noindent{\bf SET}
\begin{description}
\item[Description] The SET instruction projects the state of qubit $n$ to $|1\rangle$.
\item[Syntax] SET $n$
\item[Arguments] $n$ is in the range $0,\dots,N-1$ where $N$ is the number of qubits.
\item[Operation] $\mathrm{SET}=|1\rangle_n\langle 1|_n$
\item[Graphical symbol]\hskip 10pt $\Qcircuit @C=1.3em @R=.4em {&\gate{1}&\qw}$
\item[Note] This instruction fails if the projection results in to a state with amplitude zero.
\end{description}
\medskip

\noindent{\bf DEPOLARIZING CHANNEL}
\begin{description}
\item[Description] Insert X, Y, or Z gates with specified probabilities to mimic gate errors.
\item[Syntax] DEPOLARIZING CHANNEL\ \ \ \verb!P_X = !$p_x$ , \verb!P_Y = !$p_y$ , \verb!P_Z = !$p_z$ , SEED = $k$.
\item[Arguments] may appear in any order and any of them is optional.
Missing arguments are assumed to have value zero.
The values of the arguments should satisfy
$0\le p_x,\;p_y,\;p_z\le1$ and $0\le p_x+p_y+p_z\le1$ and $k$ must be a number smaller than $2^{31}-1$.
If $k$ is zero or negative, \QC\ takes the value provided by the operating system
as the seed for the random number generator.
\item[Operation] After each gate operation, \QC\ performs
an X gate on each qubit with probability $p_x$,
a Y gate on each qubit with probability $p_y$,
and
a Z gate on each qubit with probability $p_z$.
\item[Graphical symbol] example for $N=4$
\hskip 10pt \Qcircuit @C=1em @R=0em {
& \multigate{3}{\mathrm{DPC}} & \qw \\
& \ghost{\mathrm{DPC}} & \qw \\
& \ghost{\mathrm{DPC}} & \qw \\
& \ghost{\mathrm{DPC}} & \qw
}
\item[Note 1] This instruction should appear after QUBITS $N$ and before the first gate instruction.
\item[Note 2] For an example see ``{\bf Example: input}''. Removing `! ' from the second line
instructs \QC\ to insert X, Y, or Z gates with specified probabilities.
\end{description}
\medskip

\noindent{\bf EXIT}
\begin{description}
\item[Description] EXIT instruction terminates execution.
\item[Syntax] EXIT
\item[Arguments] None.
\item[Operation] The EXIT instruction forces \QC\ to measure all qubits and terminate.
It can appear at any point in the instruction list. It is useful for debugging.
\end{description}
\medskip

\section{Illustrative example}\label{APPc}

\begin{figure*}[ht]
\begin{center}
\mbox{
    \Qcircuit @C=0.8em @R=.8em {
      &\lstick{\KET{0}}&\gate{H}&\gate{T}&\gate{H}&\ctrl{1}&\ctrl{2}&\gate{X}&\ctrl{3}&\qw     &\ctrl{4}&\qw     &\qw     &\qw     &\targ    &\qw     &\qw      &\qw     &\qw      &\qw     &\qw     &\qw     &\qw\\
      &\lstick{\KET{0}}&\qw     &\qw     &\qw     &\targ   &\qw     &\qw     &\qw     &\ctrl{2}&\qw     &\qw     &\qw     &\qw     &\qw      &\qw     &\targ    &\qw     &\qw      &\qw     &\qw     &\qw     &\qw\\
      &\lstick{\KET{0}}&\qw     &\qw     &\qw     &\qw     &\targ   &\qw     &\qw     &\qw     &\qw     &\ctrl{2}&\qw     &\qw     &\qw      &\qw     &\qw      &\qw     &\targ    &\qw     &\qw     &\qw     &\qw\\
      &\lstick{\KET{0}}&\qw     &\qw     &\qw     &\qw     &\qw     &\qw     &\targ   &\targ   &\qw     &\qw     &\gate{M}&\qw     &\ctrl{-3}&\qw     &\ctrl{-2}&\gate{X}&\ctrl{-1}&\gate{X}&\gate{H}&\gate{0}&\qw\\
      &\lstick{\KET{0}}&\qw     &\qw     &\qw     &\qw     &\qw     &\qw     &\qw     &\qw     &\targ   &\targ   &\qw     &\gate{M}&\ctrl{-4}&\gate{X}&\ctrl{-3}&\gate{X}&\ctrl{-2}&\qw     &\gate{H}&\gate{1}&\qw\\
      }
}
  \caption{\label{fig2}Quantum circuit performing error correction on the top three qubits. The
  corresponding \QC\ input file is  listed in Example: input.
  Qubits are numbered from zero (top) to four (bottom).
  Reading from left to right, the first three gates prepare the initial state, the next two (CNOT) gates perform the encoding,
  the $X$ gate on qubit 0 introduces a spin flip error, the next 11 gates detect and correct the error and  the last 3 (2)
  gates on qubit 3 (4) illustrate how to reset a qubit to 0 (1).
   }
\end{center}
\end{figure*}
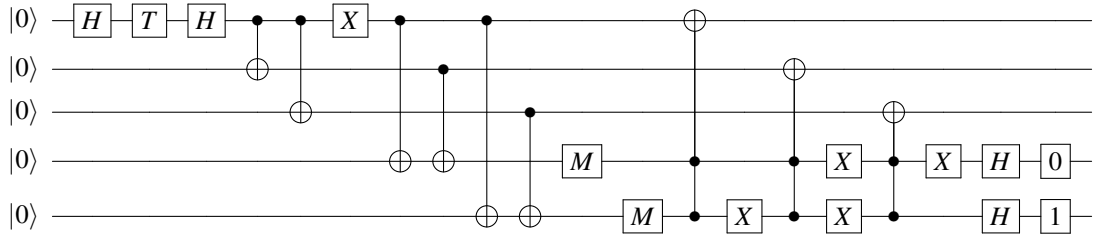

In this section, we give a simple example that shows how \QC\ processes the input file with assembler-like instructions representing
the quantum-gate circuit, in Fig.~\ref{fig2} we show a circuit that uses qubits 3 and 4 to perform
error-correction~\cite{NIEL10,DEVI13} on the logical qubit encoded in the qubits 0, 1, and 2.
The list of \QC\ instructions is given in box ``{\bf Example: input}''.
For the definition of the mnemonics see \ref{APPb}.

\smallskip
\framebox{
\parbox[ht]{0.85\hsize}{
\noindent{\bf Example: input}
\begin{lstlisting}[language=Fortran,stepnumber=1,numbers=left,basicstyle=\footnotesize]
QUBITS         5
! DEPOLARIZING CHANNEL p_X =0.01, p_Y =0.01

H 0 ! initial state
T 0
H 0 ! initial state

CNOT 0 1 ! encode
CNOT 0 2
BEGIN MEASUREMENT

!x 1 ! error
x 0 ! error

CNOT 0 3 ! correct
CNOT 1 3
CNOT 0 4
CNOT 2 4
M 3
M 4
TOFFOLI 3 4 0
X 4
TOFFOLI 3 4 1
X 3
X 4
TOFFOLI 3 4 2
X 3

H 3
CLEAR 3  ! must be preceeded by Hadamard
H 4
SET 4 ! must be preceeded by Hadamard

BEGIN MEASUREMENT
GENERATE EVENTS 8192 -1
\end{lstlisting}
}}
\medskip

Lines 4 to 6 prepare (starting from the state with all qubits in state $|0\rangle$) a nontrivial input state to the
error-correction circuit.
Lines 8 and 9 encode the two physical qubits into logical ones.
Line 10 instructs \QC\ to measure and print out the expectation values of the three components of the qubit (i.e. the expectation
values of the Pauli matrices).
Line 13 introduces a single-qubit error.
Lines 19 and 20 instruct \QC\ to perform a projective measurement on qubits 3 and 4, respectively.
Lines 15 to 27 are the instructions to detect and correct the error, if any.
Lines 29 (31) and 30 (32) show the instructions to prepare qubit 3 (4), which has undergone a projective measurement in line 19 (20),
for later re-use.
Line 35 instructs \QC\ to generate an output file with 8192 events, meaning bit strings representing basis states, sampled from the final state
of the quantum computer.

The relevant parts of the output, i.e. the expectation values of the
three components of the five qubits in the initial and final state,
when \QC\ is run with the file ``{\bf Example: input}'' as input,
(i) with line 13 commented out (no single-qubit error),
(ii) with one single-qubit error on qubit 0,
and (iii) with the `!' in line 12 removed (errors on qubits 0 and 1),
are shown in
``{\bf Example: output (i)}'',
``{\bf Example: output (ii)}'',
and
``{\bf Example: output (iii)}'', respectively.
Box ``{\bf Example: output (ii)}'' demonstrates that the error-correction code
indeed detects and corrects a single-qubit error while
``{\bf Example: output (iii)}'' shows that it fails to correct two-qubit errors.

\smallskip
\framebox{
\parbox[ht]{0.85\hsize}{
\noindent{\bf Example: output (i)}
\begin{lstlisting}[language=Fortran,stepnumber=1,numbers=left,basicstyle=\footnotesize]
---i---<Qx(i)>----<Qy(i)>----<Qz(i)>-
   0  0.500E+00  0.500E+00  0.146E+00
   1  0.500E+00  0.500E+00  0.146E+00
   2  0.500E+00  0.500E+00  0.146E+00
   3  0.500E+00  0.500E+00  0.000E+00
   4  0.500E+00  0.500E+00  0.000E+00
-------------------------------------

---i---<Qx(i)>----<Qy(i)>----<Qz(i)>-
   0  0.500E+00  0.500E+00  0.146E+00
   1  0.500E+00  0.500E+00  0.146E+00
   2  0.500E+00  0.500E+00  0.146E+00
   3  0.500E+00  0.500E+00  0.000E+00
   4  0.500E+00  0.500E+00  0.100E+01
-------------------------------------
\end{lstlisting}
}}
\medskip

\smallskip
\framebox{
\parbox[ht]{0.85\hsize}{
\noindent{\bf Example: output (ii)}
\begin{lstlisting}[language=Fortran,stepnumber=1,numbers=left,basicstyle=\footnotesize]
---i---<Qx(i)>----<Qy(i)>----<Qz(i)>-
   0  0.500E+00  0.500E+00  0.146E+00
   1  0.500E+00  0.500E+00  0.146E+00
   2  0.500E+00  0.500E+00  0.146E+00
   3  0.500E+00  0.500E+00  0.000E+00
   4  0.500E+00  0.500E+00  0.000E+00
-------------------------------------

---i---<Qx(i)>----<Qy(i)>----<Qz(i)>-
   0  0.500E+00  0.500E+00  0.146E+00
   1  0.500E+00  0.500E+00  0.146E+00
   2  0.500E+00  0.500E+00  0.146E+00
   3  0.500E+00  0.500E+00  0.000E+00
   4  0.500E+00  0.500E+00  0.100E+01
-------------------------------------
\end{lstlisting}
}}
\medskip

\smallskip
\framebox{
\parbox[ht]{0.85\hsize}{
\noindent{\bf Example: output (iii)}
\begin{lstlisting}[language=Fortran,stepnumber=1,numbers=left,basicstyle=\footnotesize]
---i---<Qx(i)>----<Qy(i)>----<Qz(i)>-
   0  0.500E+00  0.500E+00  0.146E+00
   1  0.500E+00  0.500E+00  0.146E+00
   2  0.500E+00  0.500E+00  0.146E+00
   3  0.500E+00  0.500E+00  0.000E+00
   4  0.500E+00  0.500E+00  0.000E+00
-------------------------------------

---i---<Qx(i)>----<Qy(i)>----<Qz(i)>-
   0  0.500E+00  0.500E+00  0.854E+00
   1  0.500E+00  0.500E+00  0.854E+00
   2  0.500E+00  0.500E+00  0.854E+00
   3  0.500E+00  0.500E+00  0.000E+00
   4  0.500E+00  0.500E+00  0.100E+01
-------------------------------------
\end{lstlisting}
}}
\medskip

\section*{References}

\bibliographystyle{elsarticle-num}
\bibliography{../../../all18,../BIBL/QCsimulators}
\end{document}